\pdfoutput=1

\RequirePackage{setspace}
\documentclass[english,runningheads,a4paper]{llncs}[2018/03/10]

\usepackage[english]{babel}

\usepackage[%
  rm={oldstyle=false,proportional=true},%
  sf={oldstyle=false,proportional=true},%
  qt=false%
]{cfr-lm}
\usepackage[scaled=.975]{inconsolata}
\usepackage[scr]{rsfso}

\usepackage[T1]{fontenc}
\usepackage[utf8]{inputenc}

\usepackage{csquotes}

\RequirePackage[%
  babel,%
  final,%
  expansion=alltext,%
  protrusion=alltext-nott]{microtype}%
\DisableLigatures{encoding = T1, family = tt*}

\usepackage[table,usenames,dvipsnames,svgnames]{xcolor}

\usepackage{graphicx,transparent,tcolorbox,wrapfig}
\usepackage{relsize,xparse,xspace,upquote,url}

\usepackage{bookmark,booktabs, hhline, multirow, tabularx}
\newcolumntype{C}{>{\centering\arraybackslash}X}

\usepackage{amsmath,amssymb,bm,centernot,mathdots,nicefrac,proof}
\usepackage{pifont,textcomp,textgreek}
\usepackage[geometry,misc]{ifsym}
\usepackage[makeroom]{cancel}
\usepackage[epsilon]{backnaur}
\newenvironment{bnfsplit}[1][0.8]
  {\hspace{-0.75em}\minipage[t]{#1\linewidth}\setstretch{1.1}$\hspace{0.85em}}
  {$\endminipage}

\usepackage[%
  square,        % for square brackets
  comma,         % use commas as separators
  numbers,       % for numerical citations;
  sort
]{natbib}

\usepackage[inline]{enumitem}
\newcommand{\inliststyle}[1]{\textbf{\smaller#1}}
\newlist{inlist}{enumerate*}{1}
\setlist[inlist]{label={\inliststyle{(\arabic*)}}}

\usepackage[hang]{footmisc}
\usepackage{stfloats}
\fnbelowfloat
\newlength{\textfloatsepsave}
\usepackage{subcaption}

\usepackage{algorithm}
\usepackage[noend]{algpseudocode}

\usepackage{listings}
\renewcommand{\lstlistingname}{List.}
\usepackage{chngcntr}
\AtBeginDocument{\counterwithout{lstlisting}{section}}

\usepackage[capitalise,nameinlink]{cleveref}
\crefname{section}{\S\!}{\S\S\!}
\Crefname{section}{Section}{Sections}
\crefname{figure}{Fig.}{Figs.}
\Crefname{figure}{Fig.}{Figs.}
\crefname{equation}{Equation}{Equations}
\Crefname{equation}{Equation}{Equations}
\crefname{listing}{\lstlistingname}{\lstlistingname}
\Crefname{listing}{Listing}{Listings}

\usepackage[all]{hypcap}

\hypersetup{colorlinks,
            citecolor=blue!90!black,
            linkcolor=red!80!black,
            urlcolor=blue!90!black}

\usepackage[group-four-digits,per-mode=fraction]{siunitx}

\usepackage{fixme}
\fxsetup{draft,inline,nomargin,theme=color}

\colorlet{success}{green!42!black}
\colorlet{failure}{red!64!black}
\colorlet{bgcolor}{white!88!black}
\colorlet{bordercolor}{white!72!black}
\colorlet{darkbordercolor}{white!64!black}
\colorlet{darkerbordercolor}{white!36!black}
\colorlet{fgcolor}{white!48!black}

\algrenewcommand\algorithmicindent{1.75em}

\algnewcommand{\IfThen}[2]{% \IfThen{<if>}{<then>}
  \State \algorithmicif\ #1\ \algorithmicthen\ #2}
\algnewcommand{\IfThenElse}[3]{% \IfThenElse{<if>}{<then>}{<else>}
  \State \algorithmicif\ #1\ \algorithmicthen\ #2\ \algorithmicelse\ #3}

\algrenewcommand\alglinenumber[1]{\smaller \textcolor{darkbordercolor}{\texttt{#1}}\hspace{0.5em}}
\algnewcommand{\LineComment}[1]{\State {\smaller\hspace{-2em}\textbf{\(\blacktriangleright\)\: #1}}}
\algrenewcommand{\algorithmiccomment}[1]{\hfill {\smaller \textbf{\(\blacktriangleright\)\: #1}}}

\algnewcommand\algorithmicthrow{\textbf{throw}}
\algnewcommand{\Throw}[1]{\State \algorithmicthrow\ #1}

\algnewcommand\algorithmicassume{\textbf{assume}}
\algnewcommand\Assume[1]{\State\algorithmicassume\ #1}
\algnewcommand\algorithmicassert{\textbf{assert}}
\algnewcommand\Assert[1]{\State\algorithmicassert\ #1}

\algblock{Func}{EndFunc}
\algnewcommand\algorithmicfunc{\textbf{func}}
\algnewcommand\algorithmicendfunc{\textbf{end\ func}}
\algrenewtext{Func}[2]{\algorithmicfunc\ \textsc{#1}\,(#2)}
\algrenewtext{EndFunc}{\algorithmicendfunc}
\algtext*{EndFunc}

\algblock{ParFor}{EndParFor}
\algnewcommand\algorithmicparfor{\textbf{parallel\,for}}
\algnewcommand\algorithmicpardo{\textbf{do}}
\algnewcommand\algorithmicendparfor{\textbf{end\ parallel\,for}}
\algrenewtext{ParFor}[1]{\algorithmicparfor\ #1\ \algorithmicpardo}
\algrenewtext{EndParFor}{\algorithmicendparfor}
\algtext*{EndParFor}

\algblock{ForEach}{EndForEach}
\algnewcommand\algorithmicforeach{\textbf{for\,each}}
\algnewcommand\algorithmicforeachdo{\textbf{do}}
\algnewcommand\algorithmicendforeach{\textbf{end\ for\,each}}
\algrenewtext{ForEach}[1]{\algorithmicforeach\ #1\ \algorithmicforeachdo}
\algrenewtext{EndForEach}{\algorithmicendforeach}
\algtext*{EndForEach}

\newcommand{\CodePunct}[1]{\textcolor{gray!80!black}{\texttt{#1}}}

\lstdefinelanguage{SyGuS}{
  alsoletter={0, 1, +, -, *, =, <, >},
  literate={(}{{\CodePunct{(}}}1
           {)}{{\CodePunct{)}}}1,
  texcl=true, % latex in comments
  morecomment=[l]{;},
  morestring=[b]",
  keywords=[1]{NIA, LIA},
  keywords=[2]{set-logic, define-fun, declare-var, check-synth, synth-fun, constraint},
  keywords=[3]{Constant, Variable},
  keywords=[4]{Int, Bool},
  keywords=[5]{0, 1, 2, 1000, true, false},
  keywords=[6]{+, -, *, =, <, >, >=, <=, div, mod},
  keywords=[7]{and, or, not, ite, =>}
}
\newcommand{\eg}{e.g.,\xspace}
\newcommand{\ie}{i.e.,\xspace}

\newcommand{\range}[2]{#1\text{\kern0.1em--\kern0.1em}#2}

\newcommand{\sygus}[1]{\text{\relscale{0.95}\lstinline[language=SyGuS]{#1}}}

\newcommand{\HE}{HE\xspace}

\newcommand{\tool}[1]{\textsc{#1}\xspace}
\newcommand{\HEAlgo}{\tool{HEnum}}
\newcommand{\LoopInvGen}{\tool{LoopInvGen}}
\newcommand{\LoopInvGenComp}{\tool{LoopInvGen\kern-0.1em$^{18}$}}
\newcommand{\HELoopInvGen}{\tool{\LoopInvGen\kern-0.35em+\kern-0.1em\HE}}
\newcommand{\CVC}{\tool{CVC4}}
\newcommand{\EUSolver}{\tool{EUSolver}}
\newcommand{\Sketch}{\tool{SketchAC}}
\newcommand{\Stoch}{\tool{Stoch}}

\newcommand{\PLearn}{\tool{PLearn}}

\newcommand{\Theory}[1]{\textsf{\relscale{0.95}#1}\xspace}
\newcommand{\NIA}{\Theory{Peano}}
\newcommand{\NLMult}{\Theory{Polynomials}}
\newcommand{\LIA}{\Theory{Polyhedra}}
\newcommand{\NoMult}{\Theory{Octagons}}
\newcommand{\NoArith}{\Theory{Intervals}}
\newcommand{\Equality}{\Theory{Equalities}}

\newcommand{\components}{\ensuremath{\mathscr{C}}}
\newcommand{\query}{\ensuremath{\textsf{S}}\xspace}
\newcommand{\totalcost}[1]{\scalebox{#1}{\texttau}}

\newcommand{\values}[1]{\ensuremath{\textnormal{\textsf{values}}(#1)}}
\newcommand{\operators}[1]{\ensuremath{\textnormal{\textsf{operators}}(#1)}}
\newcommand{\basecomps}[1]{\ensuremath{\textnormal{\textsf{components}}(#1)}}

\NewDocumentCommand
  {\io}
  { O{} }
  {\ensuremath{\langle x_{#1}, y_{#1} \rangle}\xspace}

\newcommand{\spec}{\ensuremath{\phi}\xspace}
\newcommand{\learner}{\ensuremath{\mathcal{L}}\xspace}
\newcommand{\theory}{\ensuremath{\mathbb{T}}\xspace}
\newcommand{\grammar}{\ensuremath{\mathcal{E}}\xspace}
\newcommand{\grammars}[1]{\ensuremath{\grammar_{1\ldots#1}}\xspace}
\newcommand{\expressions}[1]{\ensuremath{e\raisebox{-0.2em}{\scalebox{0.65}{\ensuremath{1\ldots#1}}}}\xspace}

\NewDocumentCommand
  {\oracle}
  { O{\spec} }
  {\ensuremath{\mathcal{O}_{#1}}\xspace}
\NewDocumentCommand
  {\sygustuple}
  { O{f} O{X} O{Y} O{\spec} O{\grammar} O{\theory} }
  {\ensuremath{\left\langle \scalebox{0.95}{\ensuremath{#1}}_\mathsmaller{\!#2 \to #3} \,|\, \scalebox{0.95}{\ensuremath{#4, #5}} \right\rangle_\mathsmaller{\,#6}}\xspace}
\newcommand{\doo}{\ensuremath{\textnormal{\textOmega}}\xspace}

\newcommand{\naturals}{\ensuremath{\mathbb{N}}\xspace}
\newcommand{\integers}{\ensuremath{\mathbb{Z}}\xspace}

\newcommand{\bools}{\ensuremath{\mathbb{B}}\xspace}

\let\oldiff\iff
\renewcommand\iff{\mathrel{\resizebox{2.75em}{0.475em}{$\;\oldiff\;$}}}
\let\oldimplies\implies
\renewcommand\implies{\mathrel{\resizebox{2.25em}{0.475em}{$\;\oldimplies\;$}}}
\let\oldsubseteq\subseteq
\renewcommand\subseteq{\mathrel{\raisebox{0.5pt}{$\oldsubseteq$}}}
\let\oldsubset\subset
\renewcommand\subset{\mathrel{\raisebox{0.5pt}{$\oldsubset$}}}
\let\oldin\in
\renewcommand\in{\mathrel{\raisebox{0.5pt}{$\oldin$}}}

\newcommand\defeq{\mathrel{\overset{\makebox[0pt]{\mbox{\normalfont\tiny\sffamily def}}}{=}}}
\newcommand\order{\mathrel{\textnormal{\larger[1.5]\ensuremath{\triangleleft}}}}

\DeclareMathOperator*{\bwedge}{\raisebox{-0.05em}{\scalebox{1.25}{$\wedge$}}}
\DeclareMathOperator*{\bvee}{\raisebox{-0.05em}{\scalebox{1.25}{$\vee$}}}

\NewDocumentCommand
  {\vapprox}
  { O{\theory} }
  {\mathrel{\raisebox{-0.155em}{\scalebox{1.5}[2]{$\shortmid$}}\mkern-4mu\raisebox{-0.01em}{$\approx$}_\mathsmaller{\,#1}}}

\newcommand{\counter}{\mathrel{\rightsquigarrow\mkern-12.5mu\scalebox{0.875}{\raisebox{0.03em}{$\times$}}}}
\NewDocumentCommand
  {\counteredby}
  { O{\spec} }
  {\ensuremath{\counter_\mathsmaller{#1}\!}}

\NewDocumentCommand
  {\maxunique}
  { O{\grammars{p}} O{j} O{k} }
  {\ensuremath{\mathcal{U}\bm{[}\scalebox{0.9}{\ensuremath{#1}}\bm{]}_{#2}^{#3}}}

\begin{document}

\title{Overfitting in Synthesis: Theory and Practice}

\author{
  Saswat Padhi\,\inst{1}\thanks{Contributed during an internship at Microsoft Research, India.}%
              \textsubscript{\kern-1em\smaller\raisebox{-0.2em}{\href{mailto:padhi@cs.ucla.edu}{\Letter}}} \and
  Todd Millstein\,\inst{1} \and
  Aditya Nori\,\inst{2} \and
  Rahul Sharma\,\inst{3}
}
\authorrunning{Saswat Padhi \and Todd Millstein \and Aditya Nori \and Rahul Sharma}

\institute{
  University of California, Los Angeles, USA \\[-0.175em]
  \email{\{padhi,todd\}@cs.ucla.edu} \and
  Microsoft Research, Cambridge, UK \\[-0.175em]
  \email{adityan@microsoft.com} \and
  Microsoft Research, Bengaluru, India \\[-0.175em]
  \email{rahsha@microsoft.com}
}

\maketitle\vspace{-1.4em}

\begin{abstract}
    \setlength{\parindent}{1.75em}
    In syntax-guided synthesis (SyGuS),
    a synthesizer's goal is to automatically generate
    a program belonging to a grammar of possible implementations that meets a logical specification.
    We investigate a common limitation across state-of-the-art SyGuS tools
    that perform counterexample-guided inductive synthesis (CEGIS).
    We empirically observe that as the expressiveness of the provided grammar increases,
    the performance of these tools degrades significantly.

    We claim that this degradation is not only due to a larger search space, but also due to \emph{overfitting}.
    We formally define this phenomenon and prove \emph{no-free-lunch} theorems for SyGuS,
    which reveal a fundamental tradeoff between synthesizer performance and grammar expressiveness.

    A standard approach to mitigate overfitting in machine learning is
    to run multiple learners with varying expressiveness in parallel.
    We demonstrate that this insight can immediately benefit existing SyGuS tools.
    We also propose a novel single-threaded technique called \emph{hybrid enumeration}
    that interleaves different grammars
    and outperforms the winner of the 2018 SyGuS competition (\texttt{Inv} track), 
    solving more problems and achieving a $5\times$ mean speedup.
\end{abstract}

\vspace{-2.1em}\section{Introduction}
\label{sec:introduction}

\noindent
% Automatic generation of programs, known as \emph{program synthesis},
% aims to increase the productivity of end users by mechanizing the process of
% constructing software that meets their intents and constraints,
% expressed as a specification for the program.
% Although program synthesis is difficult in general, a number of recent
% works~\citep{thesis/solar-lezama/sketching,oopsla15/polozov/flashmeta,
% asplos13/schkufza/stoke,popl16/bornholt/metasketch}
% demonstrate that restricting the search space using a syntactic template
% for programs allows for efficient synthesis strategies.
The \emph{syntax-guided synthesis} (SyGuS) framework~\citep{fmcad13/alur/sygus}
provides a unified format to describe a program synthesis problem by supplying
\begin{inlist}
      \item a logical specification for the desired functionality, and
      \item a grammar of allowed implementations.
\end{inlist}
Given these two inputs, a SyGuS tool searches through the programs that are permitted by the grammar
to generate one that meets the specification.
Today, SyGuS is at the core of several state-of-the-art program
synthesizers~\citep{pacmpl18/ezudheen/horn-ice,pldi16/padhi/data,
tacas17/alur/scaling,fse17/le/s3,pldi18/lee/euphony},
many of which compete annually in the SyGuS competition~\citep{web/sygus-comp, corr18/alur/sygus-comp}.

% However, the performance of existing synthesizers is very sensitive to the choice of grammars.
% The grammar must be expressive enough to permit the kinds of programs
% required to solve a synthesis task.
% Otherwise, the synthesizer would fail to find a solution.
%However, 
% As grammar expressiveness increases, so does the size of the search space of candidate programs.
We demonstrate empirically that five state-of-the-art SyGuS tools
are very sensitive to the choice of grammar.
% Specifically, as the grammar expressiveness increases,
% the number of synthesis problems that can be solved by these tools
% (within a generous timeout period) degrades significantly.
%and this trend holds for four state-of-the-art SyGuS tools that we evaluated.
Increasing grammar expressiveness allows the tools to solve some problems
that are unsolvable with less-expressive grammars.
However, it also causes them to fail on many problems that
the tools are able to solve with a less expressive grammar.
% This means that in practice users have the burden of identifying a ``just-right'' grammar for a particular class of synthesis problems in order to use these tools effectively, which is hard in general~\citep{cav18/abate/cegis-t}.  
We analyze the latter behavior both theoretically and empirically
and present techniques that make existing tools much more robust
in the face of increasing grammar expressiveness.

We restrict our investigation to a widely used approach~\citep{cacm18/alur/search} to SyGuS
called \emph{counterexample-guided inductive synthesis} (CEGIS)~\citep[\S 5]{sttt13/solar-lezama/program}.
In this approach, the synthesizer is composed of a learner and an oracle.
The learner iteratively identifies a candidate program that is consistent
with a given set of examples (initially empty) and queries the oracle to
either prove that the program is \emph{correct}, \ie
meets the given specification, or obtain a counterexample
that demonstrates that the program does not meet the specification.
The counterexample is added to the set of examples for the next iteration.
The iterations continue until a correct program is found
or resource/time budgets are exhausted.

\vspace{-0.45em}\paragraph{Overfitting.}
To better understand the observed performance degradation,
we instrumented one of these SyGuS tools (\cref{sec:motivation.evidence-for-overfitting}).
We empirically observe that for a large number of problems,
the performance degradation on increasing grammar expressiveness
is often accompanied by a significant increase in the number of counterexamples required.
Intuitively, as grammar expressiveness increases so does the number of \emph{spurious} candidate programs,
which satisfy a given set of examples but violate the specification.
If the learner picks such a candidate, then the oracle generates a counterexample,
the learner searches again, and so on.
% Given many equally likely alternatives, the %synthesis tool can get confused and choose an %incorrect candidate.
%We formally prove that all synthesis tools are %guaranteed to suffer from this problem.
% in a \emph{no-free-lunch} theorem.
%  \fxerror{
%      May be show an example at this point.
%      The typical ML examples for overfitting won't % work because there's no concept of ``noise'' %here.
%} We demonstrate empirically that this is the %primary source of performance degradation:  as %grammar expressiveness increases, SyGuS tools %require an increasing number of CEGIS rounds in %order to disambiguate among candidates and %ultimately find one that satisfies the given %specification.

In other words, increasing grammar expressiveness increases the chances for \emph{overfitting},
a well-known phenomenon in machine learning (ML).
Overfitting occurs when a learned function explains a given set of observations
but does not generalize correctly beyond it.
Since SyGuS is indeed a form of function learning,
it is perhaps not surprising that it is prone to overfitting.
However, we identify its specific source in the context of SyGuS ---
the spurious candidates induced by increasing grammar expressiveness ---
and show that it is a significant problem in practice.
We formally define the \emph{potential for overfitting}\;($\doo$), in \cref{defn:overfitting},
which captures the number of spurious candidates.

\vspace{-0.45em}\paragraph{No Free Lunch.}
In the ML community, this tradeoff between expressiveness and overfitting
has been formalized for various settings as \emph{no-free-lunch} (NFL) theorems~\citep[\S 5.1]{book/shalev/understanding}.
Intuitively such a theorem says that for every learner there exists a function
that cannot be efficiently learned,
where efficiency is defined by the number of examples required.
We have proven corresponding NFL theorems for the CEGIS-based SyGuS setting
(\cref{thm:nfl-cegis-infinite,thm:nfl-cegis-finite}).

A key difference between the ML and SyGuS settings is the notion of \emph{$m$-learnability}.
In the ML setting, the learned function may differ from the true function,
as long as this difference (expressed as an error probability) is relatively small.
However, because the learner is allowed to make errors,
it is in turn required to learn given an arbitrary set of $m$ examples (drawn from some distribution).
In contrast, the SyGuS learning setting is \emph{all-or-nothing} ---
either the tool synthesizes a program that meets the given specification or it fails.
Therefore, it would be overly strong to require the learner to handle an arbitrary set of examples.

Instead, we define a much weaker notion of $m$-learnability for SyGuS,
which only requires that there \emph{exist} a set of $m$ examples for which the learner succeeds.
Yet, our NFL theorem shows that even this weak notion of learnability can always be thwarted:
given an integer $m \geq 0$ and an expressive enough (as a function of $m$) grammar,
for every learner there exists a SyGuS problem that cannot be learned without access to more than $m$ examples.
We also prove that overfitting is inevitable with an expressive enough grammar (\cref{thm:overfitting-finite,thm:overfitting-infinite})
and that the potential for overfitting increases with grammar expressiveness (\cref{thm:overfitting-expressiveness}).

\vspace{-0.35em}\paragraph{Mitigating Overfitting.}
Inspired by \emph{ensemble methods}~\citep{mcs00/dietterich/ensemble} in ML,
which aggregate results from multiple learners to combat overfitting (and underfitting),
we propose \PLearn\ --- a black-box framework that runs multiple parallel instances of a SyGuS tool with different grammars.
Although prior SyGuS tools run multiple instances of learners with different random seeds~\cite{cav15/jeon/adaptive,cav11/barrett/cvc4},
to our knowledge, this is the first proposal to explore multiple grammars as a means to improve the performance of SyGuS.
Our experiments indicate that \PLearn significantly improves the performance of five state-of-the-art SyGuS tools ---
\CVC~\citep{cav15/reynolds/counterexample,cav11/barrett/cvc4}, \EUSolver~\citep{tacas17/alur/scaling},
\LoopInvGen~\citep{pldi16/padhi/data}, \Sketch~\citep{cav15/jeon/adaptive, sttt13/solar-lezama/program},
and \Stoch~\citep[\small III\,F]{fmcad13/alur/sygus}.
% More generally, we believe that running multiple learners with different subsets of the input grammar
% (with varying expressiveness) can benefit all synthesizers.

However, running parallel instances of a synthesizer is computationally expensive.
Hence, we also devise a white-box approach, called \emph{hybrid enumeration},
that extends the enumerative synthesis technique~\citep{cav13/albarghouthi/escher}
to efficiently interleave exploration of multiple grammars in a single SyGuS instance.
%Today, many enumerative SyGuS solvers explore %expressions within the given grammar
%in order of increasing expression size.
%The idea is that smaller expressions are more %natural
%and are also more likely to generalize from the %given examples.
%We present a novel enumeration strategy, called %\emph{hybrid enumeration} (\HEAlgo),
%which adds a second dimension in this search %space --- a stratification of the grammar.
%Over a grammar $G$, we assume $n$ %subgrammars, called \emph{levels} henceforth,
%$G_1,\ldots,G_n$ with increasing expressiveness,
%\ie each $G_i \subset G_{i+1}$ and $G_n = G$.
%For the domain of integers, we have $5$ levels as %mentioned above.
%We propose a metric that guides the search %through this 2D space of size $\times$ level.
%At any point during the search, our metric decides %whether to search
%\emph{deeper} (greater size) or \emph{wider} %(greater level).
%Inspired by our theoretical results, our metric %estimates the expressiveness at a given level and %size
%and directs hybrid enumeration to incrementally %increase the expressiveness of the space of %expressions.
%\fxerror{
%      The 1D search with increasing size also %incrementally increases expressiveness.
%      So  why does it not work? Or do we need to %mention that ours is more \emph{fine-grained}?
%}
%
% We demonstrate empirically that for state-of-the-art SyGuS learners,
% hybrid enumeration can match the performance of \PLearn.
We implement hybrid enumeration within \LoopInvGen\footnote{%
  Our implementation is available at \url{https://github.com/SaswatPadhi/LoopInvGen}.
}
and show that the resulting single-threaded learner, \HELoopInvGen,
has negligible overhead but achieves performance comparable to that of \PLearn for \LoopInvGen.
Moreover, \HELoopInvGen significantly outperforms the winner~\citep{corr18/padhi/loopinvgen}
of the invariant-synthesis (\texttt{Inv}) track of 2018 SyGuS competition~\citep{corr18/alur/sygus-comp} ---
a variant of \LoopInvGen specifically tuned for the competition ---
including a $5\times$ mean speedup and solving two SyGuS problems that no tool in the competition could solve.

\vspace{-0.35em}\paragraph{Contributions.}
In summary, we present the following contributions:
\begin{itemize}[leftmargin=2.5em,itemsep=0.125em,topsep=0.125em]
  \item [(\cref{sec:motivation})]
        We empirically observe that, in many cases, increasing grammar expressiveness
        degrades performance of existing SyGuS tools due to \emph{overfitting}.
  \item [(\cref{sec:formalization})]
        We formally define overfitting and prove \emph{no-free-lunch} theorems for the SyGuS setting,
        which indicate that overfitting with increasing grammar expressiveness is a fundamental characteristic of SyGuS.
  \item [(\cref{sec:mitigation})]
        We propose two mitigation strategies --
        \begin{inlist}
              \item a black-box technique that runs multiple parallel instances of a synthesizer,
                    each with a different grammar, and
              \item a single-threaded enumerative technique, called \emph{hybrid enumeration},
                    that interleaves exploration of multiple grammars.
        \end{inlist}
  \item [(\cref{sec:evaluation})]
        We show that incorporating these mitigating measures in existing tools
        significantly improves their performance.
\end{itemize}

\section{Motivation}
\label{sec:motivation}

In this section, we first present empirical evidence that existing SyGuS tools
are sensitive to changes in grammar expressiveness.
Specifically, we demonstrate that as we increase the expressiveness of the provided grammar,
every tool starts failing on some benchmarks that it was able to solve with less-expressive grammars.
We then investigate one of these tools in detail.

\subsection{Grammar Sensitivity of SyGuS Tools}
\label{sec:motivation.grammar-sensitivity}

% A SyGuS problem is specified by two kinds of constraints:
% \begin{inlist}
%     \item a logical specification for the target function, and
%     \item a grammar of allowed implementations.
% \end{inlist}
% A solution, called a \emph{satisfying expression},
% for a SyGuS problem is an expression drawn from the given grammar
% such that it also satisfies the given specification.

We evaluated $5$ state-of-the-art SyGuS tools that use very different techniques:
\begin{itemize}[itemsep=0.075em,topsep=0.1em,leftmargin=1.5em]
    \item \Sketch~\citep{cav15/jeon/adaptive} extends
          the \tool{Sketch} synthesis system~\citep{sttt13/solar-lezama/program}
          by combining both explicit and symbolic search techniques.
    \item \Stoch~\citep[\small III\,F]{fmcad13/alur/sygus} performs a stochastic search for solutions.
    \item \EUSolver~\citep{tacas17/alur/scaling} combines enumeration with unification strategies.
    \item \citet{cav15/reynolds/counterexample} extend CVC4~\citep{cav11/barrett/cvc4}
          with a refutation-based approach.
    \item \LoopInvGen~\citep{pldi16/padhi/data} combines enumeration and Boolean function learning.
\end{itemize}\vspace{0.1em}

\begin{wrapfigure}[19]{r}{0.41\textwidth}
    \vspace{-3em}\hspace{-0.75em}\scalebox{0.8}{%
    \begin{minipage}{\textwidth}
    \begin{bnf*}
        \bnfprod {$\bm{b}$}
                 {\begin{bnfsplit}
                      \sygus{true} \bnfor \sygus{false} \bnfor \bnfpn{\sygus{Bool} variables} \\[0.05em]
                  \bnfor
                      \sygus{(} \sygus{not} \bnfsp \,\bm{b} \sygus{)}
                  \bnfor
                      \sygus{(} \sygus{or} \bnfsp \,\bm{b} \bnfsp \,\bm{b} \sygus{)}
                  \bnfor
                      \sygus{(} \sygus{and} \bnfsp \,\bm{b} \bnfsp \,\bm{b} \sygus{)}
                  \end{bnfsplit}} \\
            \bnfprod {$\bm{i}$}
                     {\bnfpn{\sygus{Int} constants} \!\bnfor\! \bnfpn{\sygus{Int} variables}}
    \end{bnf*}\\[-1.7em]
    \textbf{\textsf{\color{darkerbordercolor}\relscale{0.95}\hspace{0.75em} $\blacktriangleright$ Additional rule in \Equality grammar\,:}}\vspace{-1em}
    \begin{bnf*}
           \bnfprod {$\bm{b}$}
                    {\begin{bnfsplit}
                         \hspace{-0.95em}\raisebox{0.55em}{\scalebox{0.425}{$\bm{+}$}}\hspace{0.575em}
                         \sygus{(} \sygus{=} \bnfsp \,\bm{i} \bnfsp \,\bm{i} \sygus{)}
                    \end{bnfsplit}}
    \end{bnf*}\\[-1.7em]
    \textbf{\textsf{\color{darkerbordercolor}\relscale{0.95}\hspace{0.75em} $\blacktriangleright$ Additional rules in \NoArith grammar\,:}}\vspace{-1em}
    \begin{bnf*}
        \bnfprod {$\bm{b}$}
                 {\begin{bnfsplit}
                      \hspace{-0.95em}\raisebox{0.55em}{\scalebox{0.425}{$\bm{+}$}}\hspace{0.575em}
                      \sygus{(} \sygus{>} \bnfsp \,\bm{i} \bnfsp \,\bm{i} \sygus{)}
                  \bnfor
                      \sygus{(} \sygus{>=} \bnfsp \,\bm{i} \bnfsp \,\bm{i} \sygus{)} \\[0.05em]
                  \bnfor
                      \sygus{(} \sygus{<} \bnfsp \,\bm{i} \bnfsp \,\bm{i} \sygus{)}
                  \bnfor
                      \sygus{(} \sygus{<=} \bnfsp \,\bm{i} \bnfsp \,\bm{i} \sygus{)}
                 \end{bnfsplit}}
    \end{bnf*}\\[-1em]
    \textbf{\textsf{\color{darkerbordercolor}\relscale{0.95}\hspace{0.75em} $\blacktriangleright$ Additional rules in \NoMult grammar\,:}}\vspace{-1em}
    \begin{bnf*}
        \bnfprod {$\bm{i}$}
                 {\begin{bnfsplit}
                      \hspace{-0.95em}\raisebox{0.55em}{\scalebox{0.425}{$\bm{+}$}}\hspace{0.575em}
                      \sygus{(} \sygus{+} \bnfsp \,\bm{i} \bnfsp \,\bm{i} \sygus{)}
                  \bnfor
                      \sygus{(} \sygus{-} \bnfsp \,\bm{i} \bnfsp \,\bm{i} \sygus{)}
                 \end{bnfsplit}}
    \end{bnf*}\\[-1.7em]
    \textbf{\textsf{\color{darkerbordercolor}\relscale{0.95}\hspace{0.75em} $\blacktriangleright$ Additional rule in \LIA grammar\,:}}\vspace{-1em}
    \begin{bnf*}
        \bnfprod {$\bm{i}$}
                 {\begin{bnfsplit}
                      \hspace{-0.95em}\raisebox{0.55em}{\scalebox{0.425}{$\bm{+}$}}\hspace{0.575em}
                      \sygus{(} \sygus{*}\raisebox{-0.25em}{\scalebox{0.65}{\textsf{S}}} \bnfsp \,\bm{i} \bnfsp \,\bm{i} \sygus{)}
                 \end{bnfsplit}}
    \end{bnf*}\\[-1.7em]
    \textbf{\textsf{\color{darkerbordercolor}\relscale{0.95}\hspace{0.75em} $\blacktriangleright$ Additional rule in \NLMult grammar\,:}}\vspace{-1em}
    \begin{bnf*}
        \bnfprod {$\bm{i}$}
                 {\begin{bnfsplit}
                      \hspace{-0.95em}\raisebox{0.55em}{\scalebox{0.425}{$\bm{+}$}}\hspace{0.575em}
                      \sygus{(} \sygus{*}\raisebox{-0.25em}{\scalebox{0.65}{\textsf{N}}} \bnfsp \,\bm{i} \bnfsp \,\bm{i} \sygus{)}
                 \end{bnfsplit}}
    \end{bnf*}\\[-1.7em]
    \textbf{\textsf{\color{darkerbordercolor}\relscale{0.95}\hspace{0.75em} $\blacktriangleright$ Additional rule in \NIA grammar\,:}}\vspace{-1em}
    \begin{bnf*}
        \bnfprod {$\bm{i}$}
                 {\begin{bnfsplit}
                      \hspace{-0.95em}\raisebox{0.55em}{\scalebox{0.425}{$\bm{+}$}}\hspace{0.575em}
                      \sygus{(} \sygus{div} \bnfsp \,\bm{i} \bnfsp \,\bm{i} \sygus{)}
                  \bnfor
                      \sygus{(} \sygus{mod} \bnfsp \,\bm{i} \bnfsp \,\bm{i} \sygus{)}
                  \end{bnfsplit}}
    \end{bnf*}
    \end{minipage}}
    \captionsetup{skip=0.75em}
    \caption{Grammars of quantifier-free predicates over integers\,\protect\footnotemark}
    \label{fig:integer-grammars}
\end{wrapfigure}

\footnotetext{We use the \,\raisebox{0.125em}{\smaller{$\bnfpo\hspace{-0.825em}\raisebox{0.55em}{\scalebox{0.425}{$\bm{+}$}}$}}\; operator
              to append new rules to previously defined nonterminals.}

We ran these five tools on $180$ invariant-synthesis benchmarks,
which we describe in \cref{sec:evaluation}.
% We focus on loop-invariant synthesis since many benchmarks, of varying difficulty, are available for this task,
% and because they lend themselves naturally to the use of standard grammars of varying expressiveness.
We ran the benchmarks with each of the six grammars of quantifier-free predicates,
which are shown in \cref{fig:integer-grammars}.
These grammars correspond to widely used abstract domains
in the analysis of integer-manipulating programs ---
\Equality, \NoArith~\citep{ldrs77/cousot/static}, \NoMult~\citep{wcre01/mine/octagon},
\LIA~\citep{popl78/cousot/polyhedra}, algebraic expressions (\NLMult)
and arbitrary integer arithmetic (\NIA)~\citep{book/peano/arithmetic}.
The $\sygus{*}\raisebox{-0.25em}{\scalebox{0.5}{\textsf{S}}}$ operator denotes scalar multiplication,
\eg $\sygus{(}\sygus{*}\raisebox{-0.25em}{\scalebox{0.5}{\textsf{S}}}\ \sygus{2 x)}$,
and $\sygus{*}\raisebox{-0.25em}{\scalebox{0.5}{\textsf{N}}}$ denotes nonlinear multiplication,
\eg $\sygus{(}\sygus{*}\raisebox{-0.25em}{\scalebox{0.5}{\textsf{N}}}\ \sygus{x y)}$.

\begin{figure}[!b]%
    \vspace{-1em}\centering%
    \includegraphics[width=0.975\linewidth]{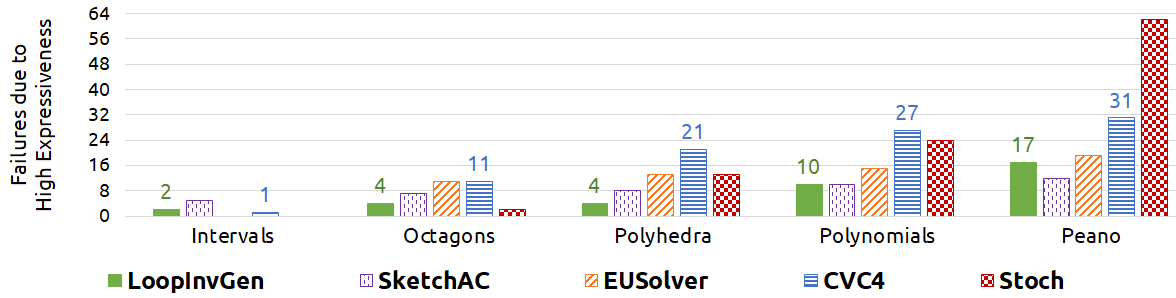}
    \captionsetup{skip=0.75em}
    \caption{For each grammar, each tool, the ordinate shows the number of benchmarks
             that \emph{fail} with the grammar but are solvable with a less-expressive grammar.\vspace{0.5em}}
    \label{fig:failures-wrt-grammars}
\end{figure}

%\fxwarning{This paragraph needs to be expanded, possibly to multiple paragraphs.  This evaluation is very cool and needs to be made clear and highlighted.  For example, walk carefully through the results for one of the tools --- how many benchmarks are solved at each grammar, how many benchmarks are solvable at some G1 but not at some superset G2, etc.  Also be clear on the experimental setup, e.g. the timeout etc.}
In \Cref{fig:failures-wrt-grammars}, we report our findings on running each benchmark on each tool with each grammar,
with a $30$-minute wall-clock timeout.
For each $\langle$tool, grammar$\rangle$ pair,
the $y$-axis shows the number of failing benchmarks that the same tool is able to solve with a less-expressive grammar.
We observe that, for each tool, the number of such failures increases with the grammar expressiveness.
For instance, introducing the scalar multiplication operator (\,$\sygus{*}\raisebox{-0.25em}{\scalebox{0.5}{\textsf{S}}}$\,)
causes \CVC to fail on $21$ benchmarks that it is able to solve
with \Equality\!($\nicefrac{4}{21}$), \NoArith\!($\nicefrac{18}{21}$), or \NoMult\!($\nicefrac{10}{21}$).
Similarly, adding nonlinear multiplication causes \LoopInvGen
to fail on $10$ benchmarks that it can solve with a less-expressive grammar.

\subsection{Evidence for Overfitting}
\label{sec:motivation.evidence-for-overfitting}

\begin{figure}[tp]%
    \centering%
    \scalebox{0.8}{%
        \def\arraystretch{0.85}%
        \setlength{\tabcolsep}{8pt}%
        \begin{tabularx}{1.25\textwidth}{C|ccc}
            \toprule
                \multicolumn{1}{C}{}
                    & \relscale{0.95}{\textbf{\textsf{Increase}}}\,($\bm{\uparrow}$)
                    & \relscale{0.95}{\textbf{\textsf{Unchanged}}}\,($\bm{=}$)
                    & \relscale{0.95}{\textbf{\textsf{Decrease}}}\,($\bm{\downarrow}$) \\
            \cmidrule{2-4}
                \relscale{0.95}{\textbf{\textsf{Expressiveness}}}$^\mathlarger{\bm{\,\uparrow}} \:\:\wedge\;$ \relscale{0.95}{\textbf{\textsf{Time}}}$^\mathlarger{\bm{\,\uparrow}} \:\rightarrow\;$ \relscale{0.95}{\textbf{\textsf{Rounds}}}$^\mathlarger{\bm{\,?}}$
                    & \relscale{1.1}{$27$\,\%}
                    & \relscale{1.1}{$67$\,\%}
                    & \relscale{1.1}{$6$\,\%} \\
                \relscale{0.95}{\textbf{\textsf{Expressiveness}}}$^\mathlarger{\bm{\,\uparrow}} \:\:\wedge\;$ \relscale{0.95}{\textbf{\textsf{Rounds}}}$^\mathlarger{\bm{\,\uparrow}} \:\rightarrow\;$ \relscale{0.95}{\textbf{\textsf{Time}}}$^\mathlarger{\bm{\,?}}$
                    & \relscale{1.1}{$79$\,\%}
                    & \relscale{1.1}{$6$\,\%}
                    & \relscale{1.1}{$15$\,\%} \\
            \bottomrule
        \end{tabularx}}
    \captionsetup{skip=0.4em}
    \caption{Observed correlation between synthesis time and number of rounds,
             upon increasing grammar expressiveness,
             with \LoopInvGen~\citep{pldi16/padhi/data} on $180$ benchmarks\vspace*{-0.35em}}
    \label{fig:loopinvgen-perf-rounds}
\end{figure}

\vspace{-0.05em}To better understand this phenomenon,
we instrumented \LoopInvGen~\citep{pldi16/padhi/data} to record the candidate expressions that it synthesizes
and the number of CEGIS iterations (called \emph{rounds} henceforth).
We compare each pair of successful runs of each of our $180$ benchmarks on distinct grammars.\footnote{%
    We ignore failing runs since they require an unknown number of rounds.
} In $65$\,\% of such pairs, we observe performance degradation with the more expressive grammar.
We also report the correlation between performance degradation and number of rounds for the more expressive grammar in each pair 
in \cref{fig:loopinvgen-perf-rounds}.

In $67$\,\% of the cases with degraded performance upon increased grammar expressiveness,
the number of rounds remains unaffected ---
indicating that this slowdown is mainly due to a larger search space.
However, there is significant evidence of performance degradation due to \emph{overfitting} as well.
We note an increase in the number of rounds for $27$\,\% of the cases with degraded performance.
Moreover, we notice performance degradation in $79$\,\% of all cases that required more rounds
on increasing grammar expressiveness.

Thus, a more expressive grammar not only increases the search space,
but also makes it more likely for \LoopInvGen to overfit --- select a spurious expression,
which the oracle rejects with a counterexample, hence requiring more rounds.
In the remainder of this section, we demonstrate this overfitting phenomenon on
the verification problem shown in \cref{fig:motivating-example},
an example by \citet{popl07/gulwani/program},
which is the \texttt{fib\_19} benchmark in the \texttt{Inv} track
of SyGuS-Comp 2018~\citep{corr18/alur/sygus-comp}.

\begin{wrapfigure}[7]{r}{0.445\textwidth}
    \vspace{-1.75em}\hspace{1.5em}\scalebox{0.8}{%
    \begin{minipage}{\textwidth}
        \begin{algorithmic}[1]
            \Assume $( 0 \leq n \;\bwedge\; 0 \leq m \leq n )$
            \Assume $( x = 0 \;\bwedge\; y = m )$
            \While {$( x < n )$}
                \State $x \gets x + 1$
                \IfThen {$( x > m )$}
                        {$y \gets y + 1$}
            \EndWhile
            \Assert $( y = n )$
        \end{algorithmic}
    \end{minipage}}
    \captionsetup{skip=0.5em}
    \caption{The \texttt{fib\_19} benchmark~\citep{popl07/gulwani/program}}
    \label{fig:motivating-example}
\end{wrapfigure}

For \cref{fig:motivating-example}, we require an inductive invariant
that is strong enough to prove that the assertion on line 6 always holds.
In the SyGuS setting, we need to synthesize a predicate $\mathcal{I} \colon \integers^{\,4} \to \bools$
defined on a symbolic state $\sigma = \langle m, n, x, y \rangle$,
that satisfies $\forall \sigma \colon \varphi(\mathcal{I}, \sigma)$
for the specification $\varphi$:\footnote{%
    We use \bools, \naturals, and \integers
    to denote the sets of all Boolean values, all natural numbers (positive integers),
    and all integers respectively.
}

{\vspace{-1em}\relscale{0.825}\begin{align*}
    \varphi(\mathcal{I}, \sigma) \ \defeq &\  \big( 0 \leq n \, \wedge \, 0 \leq m \leq n \, \wedge \, x = 0 \, \wedge \, y = m \big) \implies \mathcal{I}(\sigma)   & \qquad & \textsf{\smaller (precondition)} \\[-0.425em]
                         \bwedge    &\  \forall \sigma' \colon \big( \mathcal{I}(\sigma) \, \wedge \, T(\sigma, \sigma') \big) \implies \mathcal{I}(\sigma') & \qquad & \textsf{\smaller (inductiveness)} \\[-0.425em]
                         \bwedge    &\  \big( x \geq n \, \wedge \, \mathcal{I}(\sigma) \big) \implies y = n                                       & \qquad & \textsf{\smaller (postcondition)}
\end{align*}\vspace{-1.65em}}

\noindent
where $\sigma' = \langle m', n', x', y' \rangle$ denotes the new state after one iteration,
and $T$ is a transition relation that describes the loop body:

{\vspace{-1.125em}\relscale{0.825}\begin{align*}
    T(\sigma, \sigma') \ \defeq &\ (x < n)\ \bwedge\ (x' = x + 1)\ \bwedge\ (m' = m)\ \bwedge\ (n' = n) \\[-0.325em]
                            \bwedge &\ \big[ \, (x' \leq m \wedge y' = y)\ \bvee\ (x' > m \wedge y' = y + 1) \, \big]
\end{align*}}%
% \indent SyGuS-IF~\citep{corr14/raghothaman/language} provides a convenient format
% to describe both the specification and the grammar together for a SyGuS problem.
% In \cref{fig:sygus-example}, we show a SyGuS-IF encoding of the \textsf{forward} problem.
% Lines $5$ -- $20$ describe the grammar that restricts the syntax for the invariant.
% Lines $37$ -- $47$ encode the specification $\varphi$
% using the full vocabulary of the background theory \sygus{NIA} (line 1).
% We observe the peformance of \LoopInvGen~\citep{pldi16/padhi/data},
% the winning SyGuS solver in the \texttt{Inv} track,
% on this benchmark with various grammars.

\begin{figure}[tp]%
    \begin{subfigure}{\textwidth}%
        \centering\scalebox{0.8}{%
            \def\arraystretch{0.8}%
            \begin{tabularx}{1.25\textwidth}{CCCCCC}
                \multicolumn{6}{>{\hsize=\dimexpr6\hsize+10\tabcolsep+2\arrayrulewidth\relax\centering\arraybackslash}X}{\relscale{0.95}{Increasing expressiveness $\rightarrow$} } \\[0.1em]
                \toprule
                    \textbf{\Equality} &
                    \textbf{\NoArith} &
                    \textbf{\NoMult} &
                    \textbf{\LIA} &
                    \textbf{\NLMult} &
                    \textbf{\NIA} \\
                \midrule
                    \relscale{0.9}$\times$ &
                    \relscale{0.9}$0.32$\,s &
                    \relscale{0.9}$2.49$\,s &
                    \relscale{0.9}$2.48$\,s &
                    \relscale{0.9}$55.3$\,s &
                    \relscale{0.9}$68.0$\,s \\
                    \relscale{0.9}\texttt{FAIL} &
                    \relscale{0.9}($19$ rounds) &
                    \relscale{0.9}($57$ rounds) &
                    \relscale{0.9}($57$ rounds) &
                    \relscale{0.9}($76$ rounds) &
                    \relscale{0.9}($88$ rounds) \\
                \bottomrule
            \end{tabularx}}
        \captionsetup{skip=0.25em}
        \caption{Synthesis time and number of CEGIS iterations (rounds) with various grammars}
        \label{fig:loopinvgen-times-rounds}
    \end{subfigure} \\[1em]
    \begin{subfigure}{0.475\textwidth}%
        \linespread{1.15}\relscale{0.75}%
        \begin{itemize}[leftmargin=2.375em, itemsep=1.25em]
            \item[\textbf{\relscale{0.85} 16:}] $(x \geq n) \vee (x + 1 < n) \vee (m \geq x \,\wedge\, m = y)$
            \item[\textbf{\relscale{0.85} 28:}] $(x = y) \vee (y + m - n = x) \vee (x + 2 < n)$
            \item[\textbf{\relscale{0.85} 57:}] $(m = y) \vee (x \geq m \wedge x \geq y)$
        \end{itemize}
        \captionsetup{skip=-0.65em}
        \caption{Sample predicates with \LIA}
        \label{fig:loopinvgen-NoMult-sample}
    \end{subfigure}\hfill%
    \begin{subfigure}{0.5\textwidth}%
        \linespread{1.1}\relscale{0.725}%
        \begin{itemize}[leftmargin=3.125em, itemsep=0.4em]
            \item[\textbf{\relscale{0.85} 16:}] $(x \geq n) \vee (x + 1 < n) \; \vee$ \\[0.1em] $(2y = n) \vee (y \, (m - 1) = m)$
            \item[\textbf{\relscale{0.85} 28:}] $(y = 1) \vee (y = 0) \vee (m < 1) \vee (x^2 y > 1)$
            \item[\textbf{\relscale{0.85} 57:}] $(x + 1 \geq n) \vee (x + 2 < n) \; \vee$ \\[0.1em] $((m - n) (x - y) = 1)$
        \end{itemize}
        \captionsetup{skip=-1em}
        \caption{Sample predicates with \NIA}
        \label{fig:loopinvgen-NIA-sample}
    \end{subfigure} \\[-0.7em]
    \begin{center}\relscale{0.85}
        Solution in both grammars: $(n \geq y) \;\bwedge\; (y \geq x) \;\bwedge\; ((m = y) \,\vee\, (x \geq m \wedge x \geq y))$
    \end{center}
    \captionsetup{skip=-0.25em}
    \caption{Performance of \LoopInvGen~\citep{pldi16/padhi/data} on the \texttt{fib\_19} benchmark (\cref{fig:motivating-example}).
             In \inliststyle{(b)} and \inliststyle{(c)}, we show predicates generated at various rounds (numbered in \textbf{bold}).}
    \label{fig:motivating-example-loopinvgen-perf}
\end{figure}

In \Cref{fig:loopinvgen-times-rounds}, we report the performance of \LoopInvGen
on \texttt{fib\_19} (\cref{fig:motivating-example}) with our six grammars (\cref{fig:integer-grammars}).
It succeeds with all but the least-expressive grammar.
However, as grammar expressiveness increases,
the number of rounds increase significantly ---
from $19$ rounds with \NoArith to $88$ rounds with \NIA.

\LoopInvGen converges to the \emph{exact same} invariant with both \LIA and \NIA
but requires $30$ more rounds in the latter case.
In \cref{fig:loopinvgen-NoMult-sample,fig:loopinvgen-NIA-sample},
we list some expressions synthesized with \LIA and \NIA respectively.
These expressions are solutions to intermediate subproblems ---
the final loop invariant is a conjunction of a subset of these expressions~\citep[\S 3.2]{pldi16/padhi/data}.
Observe that the expressions generated with the \NIA grammar are quite complex and unlikely to generalize well.
\NIA's extra expressiveness leads to more spurious candidates,
increasing the chances of overfitting and making the benchmark harder to solve.

\section{SyGuS Overfitting in Theory}
\label{sec:formalization}

\vspace{-0.25em}In this section, first we formalize the \emph{counterexample-guided inductive synthesis} (CEGIS) approach~\citep{sttt13/solar-lezama/program} to SyGuS,
in which examples are iteratively provided by a verification oracle.
We then state and prove \emph{no-free-lunch} theorems,
which show that there can be no optimal learner for this learning scheme.
Finally, we formalize a natural notion of \emph{overfitting} for SyGuS
and prove that the potential for overfitting increases with grammar expressiveness.

\vspace{-0.375em}\subsection{Preliminaries}
\label{sec:formalization.learning}

\vspace{-0.125em}We borrow the formal definition of a SyGuS problem from prior work~\citep{fmcad13/alur/sygus}:
\vspace{-0.125em}\begin{definition}[SyGuS Problem]%
    Given a background theory \theory, a function symbol $f \colon X \to Y$, and constraints on $f$:
    \begin{inlist}
        \item a semantic constraint, also called a \emph{specification},
              $\spec(f, x)$ over the vocabulary of \theory along with $f$ and a symbolic input $x$, and
        \item a syntactic constraint, also called a \emph{grammar},
              given by a (possibly infinite) set \grammar of expressions over the vocabulary of the theory \theory
    \end{inlist};
    find an expression $e \in \grammar$ such that the formula
    $\forall x \in X \colon \spec(e, x)$ is valid modulo \theory.

    We denote this SyGuS problem as \sygustuple and say that it is \emph{satisfiable}
    iff there exists such an expression $e$, \ie $\exists\, e \in \grammar \colon \forall x \in X \colon \spec(e, x)$.
    We call $e$ a \emph{satisfying expression} for this problem,
    denoted as $e \models \sygustuple$.
\end{definition}

Recall, we focus on a common class of SyGuS learners, namely those that learn from examples.
First we define the notion of input-output (IO) examples that are consistent with a SyGuS specification:

\vspace{-0.25em}\begin{definition}[Input-Output Example]%
    Given a specification \spec defined on $f \colon X \to Y$ over a background theory \theory,
    we call a pair $\io \in X \times Y$ an input-output (IO) example for \spec,
    denoted as $\io \vapprox \spec$
    iff it is satisfied by some valid interpretation of $f$ within \theory, \ie
    \begin{center}\vspace{-0.375em}$
        \io \vapprox \spec \;\:\defeq\:\;
            \exists\, e_\ast \in \theory \colon e_\ast(x) = y
            \;\bwedge\;
            \big(\forall x \in X \colon \spec(e_\ast,x)\big)
    $\vspace{-0.25em}\end{center}
    % Note that $\hat{e}$ is not constrained by any grammar but only by \theory,
    % and that there could be multiple candidates for $\hat{e}$ within \theory.
\end{definition}
% We say that an expression $e$ within a grammar \grammar is consistent with a set $Z$ of IO examples,
% denoted as $Z \observedby e$, if $\forall \io \in Z \colon e(x) = y$.

% Next we define a {\em non-interactive learner},
% which is given a set of IO examples and attempts to solve a SyGuS problem in one shot:

% \begin{definition}[Non-Interactive Learner]
%     A non-interactive learner $\learner(\grammar, Z)$ is a partial function
%     that given a set \grammar of expressions and a set $Z \subseteq X \times Y$ of IO examples;
%     either fails with $\bot$ or generates an expression $e \in \grammar$ 
%     that realizes the provided examples, \ie $Z \observedby e$.
% \end{definition}

% More commonly, data-driven SyGuS learners employ a CEGIS approach,
% in which the learner interacts with a verification oracle
% to iteratively refine the candidate programs that it produces.
The next two definitions respectively formalize the two key components of a CEGIS-based SyGuS tool:
the verification oracle and the learner.

\vspace{-0.25em}\begin{definition}[Verification Oracle]%
    Given a specification \spec defined on a function $f \colon X \to Y$ over theory \theory,
    a verification oracle \oracle is a partial function that given an expression $e$,
    either returns $\bot$ indicating $\forall x \in X \colon \spec(e, x)$ holds,
    or gives a counterexample $\langle x, y\rangle$ against $e$,
    denoted as $e \counteredby \io$, such that
    \begin{center}\vspace{-0.375em}$
        e \counteredby \io \;\:\defeq\:\;
            \neg\, \spec(e,x)
            \;\bwedge\;
            e(x) \neq y
            \,\bwedge\;
            \io \vapprox \spec
    $\vspace{-0.25em}\end{center}
\end{definition}
We omit $\spec$ from the notations \oracle and $\counteredby$ when it is clear from the context.

\vspace{-0.25em}\begin{definition}[CEGIS-based Learner]%
    A CEGIS-based learner $\learner^{\oracle[]}\!(q, \grammar)$ is a partial function that
    given an integer $q \geq 0$, a set \grammar of expressions,
    and access to an oracle \oracle[] for a specification \spec defined on $f \colon X \to Y$,
    queries \oracle[] at most $q$ times and 
    either fails with $\bot$ or generates an expression $e \in \grammar$. 
    The \emph{trace}
    \begin{center}\vspace{-0.375em}$
        \big[ e_0 \counteredby[] \io[0],\: \ldots,\: e_{p-1} \counteredby[] \io[p-1],\: e_p \big] \qquad \textnormal{where } 0 \leq p \leq q
    $\vspace{-0.125em}\end{center}
    summarizes the interaction between the oracle and the learner.
    Each $e_i$ denotes the $i^\text{th}$ candidate for $f$
    and $\langle x_i, y_i \rangle$ is a counterexample $e_i$, \ie
    \begin{center}\vspace{-0.375em}$
        \big(\forall j < i \colon e_i(x_j) = y_j \wedge \spec(e_i, x_j)\big) \;\bwedge\; \big( e_i \counteredby \langle x_i, y_i \rangle \big)
    $\vspace{-0.125em}\end{center}
\end{definition}

Note that we have defined oracles and learners as (partial) functions, and hence as \emph{deterministic}.
In practice, many SyGuS tools are deterministic and this assumption simplifies the subsequent theorems.
However, we expect that these theorems can be appropriately generalized to randomized oracles and learners.
\vspace{-0.375em}\subsection{Learnability and No Free Lunch}
\label{sec:formalization.no-free-lunch}

\vspace{-0.125em}In the machine learning (ML) community,
the limits of learning have been formalized for various settings as \emph{no-free-lunch} theorems~\citep[\S 5.1]{book/shalev/understanding}.
Here, we provide a natural form of such theorems for CEGIS-based SyGuS learning.

In SyGuS, the learned function must conform to the given grammar,
which may not be fully expressive.
Therefore we first formalize grammar expressiveness:

\vspace{-0.25em}\begin{definition}[$\bm{k}$-Expressiveness]\label{def:expressiveness}%
    Given a domain $X$ and range $Y$,
    a grammar \grammar is said to be $k$-expressive
    iff \grammar can express exactly $k$ distinct $X \to Y$ functions.
\end{definition}

A key difference from the ML setting is our notion of \emph{$m$-learnability},
which formalizes the number of examples that a learner requires in order to learn a desired function.
In the ML setting, a function is considered to $m$-learnable by a learner
if it can be learned using an \emph{arbitrary} set of $m$ i.i.d. examples (drawn from some distribution).
This makes sense in the ML setting since the learned function is allowed to make errors
(up to some given bound on the error probability),
but it is much too strong for the \emph{all-or-nothing} SyGuS setting.

Instead, we define a much weaker notion of $m$-learnability for CEGIS-based SyGuS,
which only requires that there \emph{exist} a set of $m$ examples
that allows the learner to succeed.
The following definition formalizes this notion.

% \begin{definition}[Non-Interactive $\bm{m}$-Learnability]
%     Given a problem $\query = \sygustuple$ and an integer $m \geq 0$,
%     we say that \query is $m$-learnable by a non-interactive learner \learner if there exists
%     a set of $m$ examples under which \learner learns a satisfying expression for \query, \ie
%     \begin{center}$
%         \exists Z \subseteq X \times Y \colon
%                 \big(|Z| = m\big) \, \bigwedge \,
%                 \big(\forall z \in Z \colon z \vapprox \spec \big) \, \bigwedge \,
%                 \learner(\grammar, Z) \models \query
%     $\end{center}
% \end{definition}

\vspace{-0.0625em}\begin{definition}[CEGIS-based $\bm{m}$-Learnability]\label{def:learnability}%
    Given a SyGuS problem $\query = \sygustuple$ and an integer $m \geq 0$,
    we say that \query is $m$-learnable by a CEGIS-based learner \learner
    iff there exists a verification oracle \oracle[]
    under which \learner can learn a satisfying expression for \query
    with at most $m$ queries to \oracle[],
    \ie $\exists\, \oracle[] \colon \learner^{\oracle[]}\!(m, \grammar) \models \query$.
\end{definition}\vspace{-0.0625em}

Finally we state and prove the no-free-lunch (NFL) theorems,
which make explicit the tradeoff between grammar expressiveness and learnability.
Intuitively, given an integer  $m$ and an expressive enough (as a function of $m$) grammar,
for every learner there exists a SyGuS problem that cannot be solved without access to at least $m+1$ examples.
This is true despite our weak notion of learnability.

Put another way, as grammar expressiveness increases,
so does the number of examples required for learning.
On one extreme, if the given grammar is $1$-expressive,
\ie can express exactly one function,
then all satisfiable SyGuS problems are $0$-learnable ---
no examples are needed because there is only one function to learn ---
but there are \emph{many} SyGuS problems that cannot be satisfied by this function.
On the other extreme, if the grammar is $|Y|^{|X|}$-expressive,
\ie can express all functions from $X$ to $Y$,
then for every learner there exists a SyGuS problem
that requires \emph{all $|X|$ examples} in order to be solved.

Below we first present the NFL theorem
for the case when the domain $X$ and range $Y$ are finite.
We then generalize to the case when these sets may be countably infinite.
The proofs of these theorems can be found in \cref{sec:appendix.nfl-proofs}.

\begin{theorem}[NFL in CEGIS-based SyGuS on Finite Sets]\label{thm:nfl-cegis-finite}%
    Let $X$ and $Y$ be two arbitrary finite sets,
    \theory be a theory that supports equality,
    \grammar be a grammar over \theory,
    and $m$ be an integer such that $0 \leq m < |X|$.
    Then, either:% \footnote{%
    %     We use $\permute{\mathlarger n}{\,\mathlarger r}$ to denote the number of \emph{permutations},
    %     \ie different ways of selecting an ordered subset of $r$ distinct elements from a set of $n$ distinct elements.}
    \begin{itemize}[itemsep=0.25em,topsep=0.25em]
        \item \grammar is not $k$-expressive for any $k > \sum_{i \,=\, 0}^{m}$ {\larger $\frac{|X|\bm{!}\:\;|Y|^i}{(|X| \,-\, i)\bm{!}}$}, or
        \item for every CEGIS-based learner \learner,
              there exists a satisfiable SyGuS problem $\query = \sygustuple$
              such that \query is not $m$-learnable by \learner.
              Moreover, there exists a different CEGIS-based learner for which \query is $m$-learnable.
    \end{itemize}
\end{theorem}

\vspace{-0.5em}\begin{theorem}[NFL in CEGIS-based SyGuS on Countably Infinite Sets]\label{thm:nfl-cegis-infinite}%
    Let $X$ be an arbitrary countably infinite set,
    $Y$ be an arbitrary finite or countably infinite set,%\footnote{%
%        A finite domain $X$ and an infinite codomain $Y$ is absurd for a function $f \colon X \to Y$.
%    }
    \theory be a theory that supports equality,
    \grammar be a grammar over \theory,
    and $m$ be an integer such that $m \geq 0$.
    Then, either:
    \begin{itemize}[itemsep=0em,topsep=0.125em]
        \item \grammar is not $k$-expressive for any $k > \aleph_0$,
              where $\aleph_0 \defeq | \naturals |$, or
        \item for every CEGIS-based learner \learner,
              there exists a satisfiable SyGuS problem $\query = \sygustuple$
              such that \query is not $m$-learnable by \learner.
              Moreover, there exists a different CEGIS-based learner for which \query is $m$-learnable.
    \end{itemize}
\end{theorem}

\vspace{-0.75em}\subsection{Overfitting}
\label{sec:formalization.overfitting}

Last, we relate the above theory to the notion of {\em overfitting} from ML.
In the context of SyGuS, overfitting can potentially occur whenever there are multiple candidate expressions
that are consistent with a given set of examples.
Some of these expressions may not generalize to satisfy the specification,
but the learner has no way to distinguish among them
(using just the given set of examples)
and so can ``guess'' incorrectly.
We formalize this idea through the following measure:

\vspace{0.125em}\begin{definition}[Potential for Overfitting]\label{defn:overfitting}%
    Given a problem $\query = \sygustuple$ and a set $Z$ of IO examples for \spec, 
    we define the potential for overfitting $\doo$ as
    the number of expressions in \grammar that are consistent with $Z$
    but do not satisfy \query, \ie
    \begin{center}$
        \doo(\query, Z) \,\defeq\, \left\{
            \begin{array}{ll}
                \left|\big\{
                    e \in \grammar
                        \;\mid\;
                            e \not\models \query
                        \;\bwedge\;
                            \forall \io \in Z \colon e(x) = y
                \big\}\right|
                & \hspace{2.4em}\forall z \in Z \colon z \vapprox \spec \\[0.375em]
                \bot \qquad \textnormal{(undefined)}
                & \hspace{2.4em}\textnormal{otherwise}
            \end{array}
        \right.
    $\end{center}
\end{definition}

Intuitively, a zero potential for overfitting means that overfitting is not possible on the given problem
with respect to the given set of examples,
because there is no spurious candidate.
A positive potential for overfitting means that overfitting is possible,
and higher values imply more spurious candidates
and hence more potential for a learner to choose the ``wrong'' expression.

The following theorems connect our notion of overfitting to the earlier NFL theorems
by showing that overfitting is inevitable with an expressive enough grammar. 
We provide their proofs in \cref{sec:appendix.overfitting-proofs}.

\vspace{0.25em}\begin{theorem}[Overfitting in SyGuS on Finite Sets]\label{thm:overfitting-finite}%
    Let $X$ and $Y$ be two arbitrary finite sets,
    $m$ be an integer such that $0 \leq m < |X|$,
    \theory be a theory that supports equality, and
    \grammar be a $k$-expressive grammar over \theory
    for some $k >$ {\larger $\frac{|X|\bm{!}\:\;|Y|^m}{m\bm{!}\;(|X|\,-\,m)\bm{!}}$}.
    Then, there exists a satisfiable SyGuS problem $\query = \sygustuple$
    such that $\doo(\query, Z) > 0$, for every set $Z$ of $m$ IO examples for \spec.
\end{theorem}

\vspace{0.125em}\begin{theorem}[Overfitting in SyGuS on Countably Infinite Sets]\label{thm:overfitting-infinite}%
    Let $X$ be an arbitrary countably infinite set,
    $Y$ be an arbitrary finite or countably infinite set,
    \theory be a theory that supports equality, and
    \grammar be a $k$-expressive grammar over \theory
    for some $k > \aleph_0$.
    Then, there exists a satisfiable SyGuS problem $\query = \sygustuple$
    such that $\doo(\query, Z) > 0$, for every set $Z$ of $m$ IO examples for \spec.
\end{theorem}\vspace{0.125em}

Finally, it is straightforward to show that as the expressiveness of the grammar provided in a SyGuS problem increases,
so does its potential for overfitting.

\vspace{0.25em}\begin{theorem}[Overfitting Increases with Expressiveness]\label{thm:overfitting-expressiveness}%
    Let $X$ and $Y$ be two arbitrary sets,
    \theory be an arbitrary theory,
    $\grammar_1$ and $\grammar_2$ be grammars over \theory
    such that $\grammar_1 \subseteq \grammar_2$,
    \spec be an arbitrary specification over \theory and a function symbol $f \colon X \to Y$,
    and $Z$ be a set of IO examples for \spec.
    Then, we have
    \begin{center}\vspace{-0.125em}$
        \doo\big(\sygustuple[f][X][Y][\spec][\grammar_1], Z\big)
        \;\leq\;
        \doo\big(\sygustuple[f][X][Y][\spec][\grammar_2], Z\big)
    $\end{center}
 \end{theorem}

\section{Mitigating Overfitting}
\label{sec:mitigation}

% Bias-variance tradeoff is critical problem in any supervised learning technique.
% To combat overfitting, the machine learning community has developed several techniques,
% such as cross validation, regularization and boosting~\cite{ml90/schapire/strength}.

\emph{Ensemble methods}~\cite{mcs00/dietterich/ensemble} in machine learning (ML) are a standard approach to reduce overfitting.
These methods aggregate predictions from several learners to make a more accurate prediction.
In this section we propose two approaches, inspired by ensemble methods in ML,
for mitigating overfitting in SyGuS.
Both are based on the key insight from \cref{sec:formalization.overfitting}
that synthesis over a subgrammar has a smaller potential for overfitting as compared to that over the original grammar.

%,
% which are inspired by this idea.
% We propose that simultaneously exploring multiple grammars for a satisfying expression,
% would significantly reduce the impact of overfitting due to any particular grammar.

\subsection{Parallel SyGuS on Multiple Grammars}
\label{sec:mitigation.plearn}

Our first idea is to run multiple parallel instances of a synthesizer
on the same SyGuS problem but with grammars of varying expressiveness.
This framework, called \PLearn, is outlined in \cref{algo:plearn}.
It accepts a synthesis tool $\mathcal{T}$, a SyGuS problem \sygustuple,
and subgrammars \grammars{p},\!\footnote{%
    We use the shorthand $\textnormal{X}_{1, \ldots, n}$ to denote the sequence $\langle \textnormal{X}_1, \ldots, \textnormal{X}_n \rangle$.
} such that $\grammar_i \subseteq \grammar$.
%\fxwarning{I think we should leave a note here on how we would pick these sets in practice?}
The \textbf{parallel for} construct creates a new thread for each iteration.
The loop in \PLearn creates $p$ copies of the SyGuS problem, each with a different grammar from \grammars{p},
and dispatches each copy to a new instance of the tool $\mathcal{T}$.
\PLearn returns the first solution found or $\bot$ if none of the synthesizer instances succeed.

\begin{algorithm}[t]%
    \vspace{0.05em}\relscale{0.85}%
    \begin{algorithmic}[1]\linespread{1.15}\selectfont%
        \Func {\PLearn} {$\mathcal{T} \colon$Synthesis Tool,\ \ $\sygustuple \colon$Problem,\ \ $\grammars{p} \colon$Subgrammars}
        \LineComment{Requires: $\forall\, \grammar_i \in \grammars{p} \colon\; \grammar_i \subseteq \grammar$}
        \ParFor {$i \gets 1, \ldots, p$}
            \State $\query_i \gets \sygustuple[f][X][Y][\spec][\grammar_i]$
            \State $e_i \gets \mathcal{T}(\query_i)$
            \IfThen {$e_i \neq \bot$} {\Return $e_i$}
        \EndParFor
        \vspace{-0.25em}\State \Return $\bot$
      \EndFunc
    \end{algorithmic}
    \caption{The \PLearn framework for SyGuS tools.}
    \label{algo:plearn}
\end{algorithm}

Since each grammar in \grammars{p} is subsumed by the original grammar \grammar,
any expression found by \PLearn is a solution to the original SyGuS problem.
Moreover, from \cref{thm:overfitting-expressiveness}
it is immediate that \PLearn indeed reduces overfitting.

\begin{theorem}[\textnormal{\PLearn} Reduces Overfitting]%
    Given a SyGuS problem $\query = \sygustuple$,
    if \textnormal{\PLearn} is instantiated with \query
    and subgrammars \grammars{p} such that $\forall\, \grammar_i \in \grammars{p} \colon \grammar_i \subseteq \grammar$,
    then for each $\query_i = \sygustuple[f][X][Y][\spec][\grammar_i]$ constructed by \textnormal{\PLearn},
    we have that $\doo(\query_i, Z) \leq \doo(\query, Z)$
    on any set $Z$ of IO examples for \spec.
\end{theorem}
% \PLearn is sound and guarantees completeness over $\grammar_1 \cup \cdots \cup \grammar_k$,
% if the underlying learner is complete for any provided grammar.
% Similarly, \PLearn is guaranteed to terminate if the learner provides a termination guarantee as well.

A key advantage of \PLearn is that it is agnostic to the synthesizer's implementation.
Therefore, existing SyGuS learners can immediately benefit from \PLearn,
as we demonstrate in \cref{sec:evaluation.plearn}.
However, running $p$ parallel SyGuS instances can be prohibitively expensive,
both computationally and memory-wise.
The problem is worsened by the fact that many existing SyGuS tools already use multiple threads,
\eg the \Sketch~\citep{cav15/jeon/adaptive} tool spawns $9$ threads.
This motivates our {\em hybrid enumeration} technique described next,
which is a novel synthesis algorithm that interleaves exploration of multiple grammars in a single thread.

\subsection{Hybrid Enumeration}
\label{sec:mitigation.hybrid}

\vspace{-0.25em}Hybrid enumeration extends the \emph{enumerative synthesis} technique,
which enumerates expressions within a given grammar in order of size and returns the first candidate
that satisfies the given examples~\citep{cav13/albarghouthi/escher}.
Our goal is to simulate the behavior of \PLearn with an enumerative synthesizer in a single thread.
% Most practical grammars, such as the ones shown in \cref{fig:integer-grammars},
% typically are fully expressive, \ie they can express all possible functions over their domain.
% Equipped with the metric \metricsymbol, we now design a novel synthesis algorithm
% that simultaneously explores multiple grammars on a single process by interleaving them
% at various expression sizes.
% The idea is to search for expressions within various size-bounded grammars ordered by \metricsymbol,
% so as to gradually increase the expressiveness (and hence the degree of overfitting)
% in the search space.
However, a straightforward interleaving of multiple \PLearn threads would be highly inefficient
because of redundancies -- enumerating the same expression (which is contained in multiple grammars) multiple times.
% Practical grammars with increasing expressiveness, such as our integer-arithmetic grammars from \cref{fig:integer-grammars},
% often have expressions that are contained in multiple grammars.
% Therefore, exploring all expressions within all grammars would enumerate such expressions multiple times and be quite wasteful.
% We first explain a criterion for the order in which to explore expressions, that permits efficient synthesis.
% We then instantiate one such order and present an algorithm
% that explores expressions across the provided grammars in this order.
Instead, we propose a technique that
\begin{inlist}
    \item enumerates each expression at most once, and
    \item reuses previously enumerated expressions to construct larger expressions.
\end{inlist}

To achieve this, we extend a widely used~\citep{cav13/albarghouthi/escher, pldi14/perelman/tds, pldi17/feng/table} synthesis strategy,
called \emph{component-based synthesis}~\citep{icse10/jha/dit},
wherein the grammar of expressions is induced by a set of components, each of which is a typed operator with a fixed arity.
For example, the grammars shown in \cref{fig:integer-grammars}
are induced by integer components (such as \sygus{1}, \sygus{+}, \sygus{mod}, \sygus{=}, etc.)
and Boolean components (such as \sygus{true}, \sygus{and}, \sygus{or}, etc.).
Below, we first formalize the grammar that is implicit in this synthesis style.

\vspace{-0.25em}\begin{definition}[Component-Based Grammar]\label{def:component-based-grammar}%
    Given a set $\,\components$ of typed components,
    we define the \emph{component-based} grammar \grammar as
    the set of all expressions formed by well-typed component application over $\components$, \ie
    \begin{center}\vspace{-0.4em}$
        \begin{array}{rcl}
            \grammar \;=\; \{\,
                c(e_1, \ldots, e_a)
            &\mid&
                (c : \tau_1 \times \cdots \times \tau_a \to \tau) \in \components
                \,\wedge\,
                \expressions{a} \subset \grammar \\[0.125em]
            &\wedge&
                e_1 : \tau_1
                \,\wedge\,
                \cdots
                \,\wedge\,
                e_a : \tau_a
            \,\}
        \end{array}
    $\vspace{-0.5em}\end{center}
    where $e : \tau$ denotes that the expression $e$ has type $\tau$.
\end{definition}\vspace{-0.3125em}
We denote the set of all components appearing in a component-based grammar \grammar as \basecomps{\grammar}.
Henceforth, we assume that \basecomps{\grammar} is known (explicitly provided by the user) for each \grammar.
We also use \values{\grammar} to denote the subset of nullary components (variables and constants) in \basecomps{\grammar},
and \operators{\grammar} to denote the remaining components with positive arities.

% \begin{definition}[Syntactic Extension]\label{def:syntactic-extension}%
%     A grammar $\grammar_2$ is called an extension of $\grammar_1$,
%     denoted as $\grammar_2 \supseteq \grammar_1$,
%     iff $\grammar_2$ contains every expression contained in $\grammar_1$.

%     We say that an extension is \emph{syntactic},
%     denoted as $\grammar_2 \sqsupseteq \grammar_1$,
%     iff every expression in $\grammar_2 \setminus \grammar_1$ uses a terminal symbol not contained in $\grammar_1$.
%     Formally,
%     \begin{center}$
%         \grammar_2 \sqsupseteq \grammar_1 \:\defeq\:
%             \grammar_2 \supseteq \grammar_1 \;\bwedge\;
%             \forall e \in \grammar_2 \setminus \grammar_1 \colon
%                 \tau(e) \not\subseteq \Sigma_1
%     $\end{center}
%     where $\tau(e)$ denotes the set of terminal symbols in $e$,
%     and $\Sigma_1$ is the alphabet of $\grammar_1$.
% \end{definition}

The closure property of component-based grammars significantly reduces the overhead
of tracking which subexpressions can be combined together to form larger expressions.
Given a SyGuS problem over a grammar \grammar,
hybrid enumeration requires a sequence \grammars{p} of grammars
such that each $\grammar_i$ is a component-based grammar and that 
$\grammar_1 \subset \cdots \subset \grammar_p \subseteq \grammar$.
Next, we explain how the subset relationship between the grammars
enables efficient enumeration of expressions.

Given grammars $\grammar_1 \subset \cdots \subset \grammar_p$,
observe that an expression of size $k$ in $\grammar_i$
may only contain subexpressions of size $\{ 1, \ldots, (k-1) \}$ belonging to \grammars{i}.
This allows us to enumerate expressions in an order such that
each subexpression $e$ is synthesized (and cached)
before any expressions that have $e$ as a subexpression.
We call an enumeration order that ensures this property a \emph{well order}.

% Next, consider an arbitrary expression of size $k$ in one of the grammars, $\grammar_i \in \grammars{p}$,
% with $\grammar_1 \sqsubseteq \cdots \sqsubseteq \grammar_p$.
% Due to \cref{def:syntactic-extension}, the subexpressions for this expression
% may only be of size $\{ 1, \ldots, (k-1) \}$ and must belong to \grammars{i}.
% If our synthesis strategy explored these expressions before exploring expressions of size $k$ within $\grammar_i$,
% it would not be necessary to synthesize, or check if we need to synthesize, any of these subexpressions.

% An exploration strategy which always satisfies this property for all sizes $k$ and grammars $\grammar_i$,
% is called a \emph{well-order} of the space of grammars $\times$ sizes.
% 
% We formally define this property below.

\vspace{-0.25em}\begin{definition}[Well Order]\label{def:well-order}%
    Given arbitrary grammars \grammars{p},
    we say that a strict partial order $\,\order$ on $\grammars{p} \times \naturals$
    is a well order iff
    \begin{center}\vspace{-0.4em}$
        \forall\, \grammar_a, \grammar_b \in \grammars{p} :\enskip
            \forall\, k_1, k_2 \in \naturals :\enskip
            [\, \grammar_a \subseteq \grammar_b \,\wedge\, k_1 < k_2 \,]
            \implies
            (\grammar_a, k_1) \order\, (\grammar_b, k_2)
    $\vspace{-0.25em}\end{center}
\end{definition}

% Now, we present a well-order measure that gradually increases the expressiveness
% of the space of expressions being searched.
% A natural measure of expressiveness of a set of expressions is the cardinality of the set.
% We consider the measure \metric{\grammar}{k} which overapproximates the cardinality of
% the of $k$-sized expressions within \grammar.
% This is roughly $|\grammar|^k$.\footnote{%
%     For a CFG over an alphabet $\Sigma$, the number of possible $k$-length strings
%     is upper bounded by $|\Sigma|^k$.}
% It turns out, this measure is well-order:

% \input{include/figures/he-intuition}

% \cref{fig:well-order} shows an example --
% the arrows indicate a linear extension of the well order on $\grammars{5} \times \{ 1, \ldots, 4 \}$.
% The shaded region shows the set of expressions of size $\{ 1, \ldots, 3 \}$ in \grammars{4},
% which are possible subexpressions for expressions of size $4$ in $\grammar_4$.
% ,
% which guarantees that the shaded region would have been enumerated before $(\grammar_4, 4)$.

Motivated by \cref{thm:overfitting-expressiveness},
our implementation of hybrid enumeration uses a particular well order
that incrementally increases the expressiveness of the space of expressions.
For a rough measure of the expressiveness (\cref{def:expressiveness}) of a pair (\grammar, $k$),
\ie the set of expressions of size $k$ in a given grammar \grammar,
we simply overapproximate the number of syntactically distinct expressions:

\begin{theorem}%
    Let \grammars{p} be component-based grammars and $\components_i = \basecomps{\grammar_i}$.
    Then, the following strict partial order $\,\order_\mathlarger{\mkern2mu\ast}$ on $\grammars{p} \times \naturals$
    is a well order
    \begin{center}$
        \forall\, \grammar_a, \grammar_b \in \grammars{p} :\enskip
            \forall\, m, n \in \naturals :\enskip
                (\grammar_a, m) \order_\mathlarger{\mkern2mu\ast} (\grammar_b, n)
                \iff
                |\mkern2mu\components_a\mkern1mu|^m < |\mkern2mu\components_b\mkern1mu|^n
    $\end{center}
\end{theorem}

We now describe the main hybrid enumeration algorithm, which is listed in \cref{algo:hybrid-enum}.
The \HEAlgo function accepts a SyGuS problem \sygustuple,
a set \grammars{p} of component-based grammars such that $\grammar_1 \subset \cdots \subset \grammar_p \subseteq \grammar$,
a well order $\order$, and an upper bound $q \geq 0$ on the size of expressions to enumerate.
In lines \range{4}{8}, we first enumerate all values and cache them as expressions of size one.
In general $C[j, k][\tau]$ contains expressions of type $\tau$ and size $k$ from $\grammar_j \setminus \grammar_{j-1}$.
In line 9 we sort (grammar, size) pairs in some total order consistent with $\order$.
Finally, in lines \range{10}{20}, we iterate over each pair $(\grammar_j, k)$ and each operator from \grammars{j}
and invoke the \textsc{Divide} procedure (\cref{algo:divide-algo})
to carefully choose the operator's argument subexpressions ensuring
\begin{inlist}
    \item \emph{correctness} -- their sizes sum up to $k-1$,
    \item \emph{efficiency} -- expressions are enumerated at most once, and
    \item \emph{completeness} -- all expressions of size $k$ in $\grammar_j$ are enumerated.
\end{inlist}

\setlength{\textfloatsep}{1em}%
\begin{algorithm}[!b]%
    \vspace{0.05em}\relscale{0.85}%
    \begin{algorithmic}[1]\linespread{1.075}\selectfont%
        \Func {\HEAlgo} {$\sygustuple \colon$Problem, $\grammars{p} \colon$Grammars, $\order \colon$WO, $q \colon$Max. Size}
        \LineComment{Requires: component-based grammars $\grammar_1 \subset \cdots \subset \grammar_p \subseteq \grammar$ and $v$ as the input variable}
        \State $C \gets \{\}$
        \For {$i \gets 1$ \textbf{to} $p$}
            \State $V \gets$ \algorithmicif\ {$i = 1$} \algorithmicthen\ {$\values{\grammar_1}$} \algorithmicelse\ {$[\, \values{\grammar_i} \setminus \values{\grammar_{i-1}} \,]$}
            \ForEach {$(e : \tau) \in V$}
                \State $C[i, 1][\tau] \gets C[i, 1][\tau] \cup \{ e \}$
                \If {$\forall x \in X \colon \spec(\lambda v.\, e, x)$}
                    \Return $\lambda v.\, e$
                \EndIf
            \EndForEach
        \EndFor
        \vspace{-0.25em}\State $R \gets \Call{Sort}{\order,\; \grammars{p} \times \{2, \ldots, q\}}$
        \For {$i \gets 1$ \textbf{to} $|\,R\,|$}
            \State $(\grammar_j, k) \gets R[i]$
            \For {$l \gets 1$ \textbf{to} $j$}
                \State $O \gets \algorithmicif\ l = 1\ \algorithmicthen\ \operators{\grammar_1}\ \algorithmicelse\ [\, \operators{\grammar_l} \setminus \operators{\grammar_{l-1}} \,]$
                \ForEach {$(o : \tau_1 \times \cdots \times \tau_a \to \tau) \in O$}
                    \State $L \gets \Call{Divide}{a,\ k-1,\ l,\ j,\ \langle\rangle}$
                    \ForEach {$\big\langle (x_1, y_1), \ldots, (x_a, y_a) \big\rangle \in L$}
                        \ForEach {$\expressions{a} \in C[x_1, y_1][\tau_1] \times \cdots \times C[x_a, y_a][\tau_a]$}
                            \State $e \gets o(e_1, \ldots, e_a)$
                            \State $C[j, k][\tau] \gets C[j, k][\tau] \cup \{ e \}$
                            \If {$\forall x \in X \colon \spec(\lambda v.\, e, x)$}
                                \Return $\lambda v.\, e$
                            \EndIf
                        \EndForEach
                    \EndForEach
                \EndForEach
            \EndFor
        \EndFor
        \vspace{-0.25em}\State \Return $\bot$
    \EndFunc
    \end{algorithmic}
    \caption{\emph{Hybrid enumeration} to combat overfitting in SyGuS}
    \label{algo:hybrid-enum}
\end{algorithm}%
\setlength{\textfloatsep}{\textfloatsepsave}

The \textsc{Divide} algorithm generates a set of locations for selecting arguments to an operator.
Each location is a pair $(x, y)$ indicating that any expression from $C[x, y][\tau]$ can be an argument,
where $\tau$ is the argument type required by the operator.
\textsc{Divide} accepts an arity $a$ for an operator $o$, a size budget $q$,
the index $l$ of the least-expressive grammar containing $o$,
the index $j$ of the least-expressive grammar that should contain the constructed expressions of the form $o(e_1,\ldots,e_a)$,
and an accumulator $\alpha$ that stores the list of argument locations.
In lines \range{7}{9}, the size budget is recursively divided among $a-1$ locations.
In each recursive step, the upper bound $(q - a + 1)$ on $v$ ensures that
we have a size budget of at least $q - (q - a + 1) = a - 1$ for the remaining $a - 1$ locations.
This results in a call tree such that the accumulator $\alpha$ at each leaf node
contains the locations from which to select the last $a-1$ arguments,
and we are left with some size budget $q \geq 1$ for the first argument $e_1$.
Finally in lines \range{4}{5}, we carefully select the locations for $e_1$
to ensure that $o(e_1,\ldots,e_a)$ has not been synthesized before ---
either $o \in \basecomps{\grammar_j}$
or at least one argument belongs to $\grammar_j \setminus \grammar_{j-1}$.
%In particular, if $o$ and $\expressions{(a-1)}$ all belong to a prior grammar (level $< j$),
%then \textsc{Divide} only selects locations from level $j$ for $e_a$.
%Finally, to select the locations for the last argument, we either:
%\begin{inlist}
%    \item use an operator from level $j$ and subexpressions from any previous level, or
%    \item use an operator from some previous level and at least one subexpression from level $j$.
%\end{inlist}

\setlength{\textfloatsep}{1em}%
\begin{algorithm}[!t]
    \vspace{0.15em}\relscale{0.85}%
    \begin{algorithmic}[1]%
        \Func {Divide} {$a \colon$Arity, $q \colon$Size, $l \colon$Op. Level, $j \colon$Expr. Level, $\alpha \colon$Accumulated Args.}
            \LineComment{Requires: $1 \leq a \leq q \;\wedge\; l \leq j$}
            \If {$a = 1$}
                \If {$l = j \;\vee\; \exists\, \langle x, y \rangle \in \alpha \colon x = j$}
                    \Return $\big\{ (1,q) \diamond \alpha, \ldots, (j,q) \diamond \alpha \big\}$
                \EndIf
                \vspace{-0.325em}\State \Return $\big\{ (j,q) \diamond \alpha \big\}$
            \EndIf
            \vspace{-0.25em}\State $L = \{\}$
            \For {$u \gets 1$ \textbf{to} $j$}
                \For {$v \gets 1$ \textbf{to} $(q - a + 1)$}
                    \State $L \gets L \;\cup\; \Call{Divide}{a-1,\ q-v,\ l,\ j,\ (u,v) \diamond \alpha}$
                \EndFor
            \EndFor
            \vspace{-0.25em}\State \Return $L$
        \EndFunc
    \end{algorithmic}
    \vspace{0.1em}
    \caption{An algorithm to divide a given size budget among subexpressions\,\protect\footnotemark}
    \label{algo:divide-algo}
\end{algorithm}%
\setlength{\textfloatsep}{\textfloatsepsave}

\footnotetext{We use $\diamond$ as the \texttt{cons} operator for sequences, \eg $x \diamond \langle y, z \rangle = \langle x, y, z \rangle$.}

We conclude this section by stating some desirable properties satisfied by \HEAlgo.
The proofs of the following theorems can be found in \cref{sec:appendix.he-proofs}.

\vspace{-0.125em}\begin{theorem}[\textnormal{\HEAlgo} is Complete up to Size $\bm{q}$]\label{thm:he-completeness}%
    Given a SyGuS problem $\query = \sygustuple$,
    let \grammars{p} be component-based grammars over theory \theory
    such that $\grammar_1 \subset \cdots \subset \grammar_p = \grammar$,
    $\order$ be a well order on $\grammars{p} \times \naturals$,
    and $q \geq 0$ be an upper bound on size of expressions.
    Then, \textnormal{\HEAlgo}$(\query, \grammars{p}, \order, q)$
    will eventually find a satisfying expression
    if there exists one with size $\leq q$.
\end{theorem}

\vspace{-0.675em}\begin{theorem}[\textnormal{\HEAlgo} is Efficient]\label{thm:he-efficiency}%
    Given a SyGuS problem $\query = \sygustuple$,
    let \grammars{p} be component-based grammars over theory \theory
    such that $\grammar_1 \subset \cdots \subset \grammar_p \subseteq \grammar$,
    $\order$ be a well order on $\grammars{p} \times \naturals$,
    and $q \geq 0$ be an upper bound on size of expressions.
    Then, \textnormal{\HEAlgo}$(\query, \grammars{p}, \order, q)$
    will enumerate each distinct expression at most once.
\end{theorem}
\vspace{-0.875em}\section{Experimental Evaluation}
\label{sec:evaluation}

\vspace{-0.3125em}In this section we empirically evaluate \PLearn and \HEAlgo.
Our evaluation uses a set of $180$ synthesis benchmarks,\footnote{%
    All benchmarks are available at \url{https://github.com/SaswatPadhi/LoopInvGen}.
} consisting of all $127$ official benchmarks
from the \texttt{Inv} track of 2018 SyGuS competition~\citep{corr18/alur/sygus-comp} augmented
with benchmarks from the 2018 Software Verification competition (SV-Comp)~\citep{tacas17/beyer/sv-comp}
and challenging verification problems proposed in prior work~\citep{chi18/bounov/gamification,icalp05/bradley/polyrank}.
All these synthesis tasks are defined over integer and Boolean values,
and we evaluate them with the six grammars described in \cref{fig:integer-grammars}.
We have omitted benchmarks from other tracks of the SyGuS competition as they either require
us to construct $\grammars{p}$ (\cref{sec:mitigation}) by hand or lack verification oracles.
All our experiments use an $8$-core Intel\,\textsuperscript{\textregistered}
Xeon\,\textsuperscript{\textregistered} E5 machine clocked at 2.30\,GHz with 32\,GB memory
running Ubuntu\,\textsuperscript{\textregistered} 18.04.

\vspace{-0.75em}\subsection{Robustness of \textnormal{\PLearn}}
\label{sec:evaluation.plearn}

\vspace{-0.25em}For five state-of-the-art SyGuS solvers --
\begin{inlist}
    \item[\textbf{\small(a)}] \LoopInvGen~\citep{pldi16/padhi/data},
    \item[\textbf{\small(b)}] \CVC~\citep{cav15/reynolds/counterexample,cav11/barrett/cvc4},
    \item[\textbf{\small(c)}] \Stoch~\citep[\small III\,F]{fmcad13/alur/sygus},
    \item[\textbf{\small(d)}] \Sketch~\citep{cav15/jeon/adaptive,tacas17/beyer/sv-comp}, and
    \item[\textbf{\small(e)}] \EUSolver~\citep{tacas17/alur/scaling} --
\end{inlist}
we have compared the performance across various grammars,
with and without the \PLearn framework (\cref{algo:plearn}).
In this framework, to solve a SyGuS problem with the $p^\text{th}$ expressiveness level from
our six integer-arithmetic grammars (see \cref{fig:integer-grammars}),
we run $p$ independent parallel instances of a SyGuS tool,
each with one of the first $p$ grammars.
For example, to solve a SyGuS problem with the \LIA grammar,
we run four instances of a solver with the \Equality, \NoArith, \NoMult and \LIA grammars.
We evaluate these runs for each tool, for each of the $180$ benchmarks
and for each of the six expressiveness levels.

\begin{figure}[!t]%
    \centering%
    \begin{subfigure}{0.333\textwidth}%
        \includegraphics[width=\linewidth]{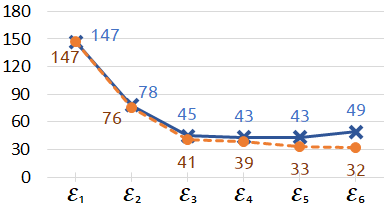}\vspace{-0.45em}
        \caption{\LoopInvGen~\citep{pldi16/padhi/data}}
    \end{subfigure}%
    \begin{subfigure}{0.333\textwidth}%
        \includegraphics[width=\linewidth]{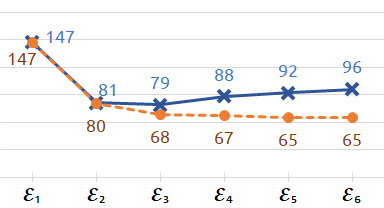}\vspace{-0.45em}
        \caption{\CVC~\citep{cav15/reynolds/counterexample, cav11/barrett/cvc4}}
    \end{subfigure}\hfill%
    \begin{subfigure}{0.275\textwidth}\vspace{0.25em}\relscale{0.725}\sf%
        \textcolor[rgb]{0.18431,0.33333,0.59216}{\bf\sf Solid blue curves (\raisebox{-0.5em}{\textbf{\Cross}})} show original failure counts. \\[0.75em]
        \textcolor[rgb]{0.74902,0.34118,0.04706}{\bf\sf Dashed orange curves ($\bullet$)} show failure counts with \textnormal{\PLearn}. \\[-0.75em]
        \begin{center} \relscale{1.1} Timeout = $30$\,min. (wall-clock) \end{center}
    \end{subfigure}\hfill%
    \begin{subfigure}{0.01\textwidth}\end{subfigure} \\[0.675em]
    \begin{subfigure}{0.333\textwidth}%
        \includegraphics[width=\linewidth]{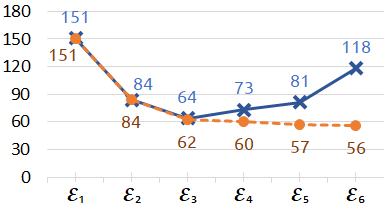}\vspace{-0.5em}
        \caption{\Stoch~\citep[\small III\,F]{fmcad13/alur/sygus}}
    \end{subfigure}%
    \begin{subfigure}{0.333\textwidth}%
        \includegraphics[width=\linewidth]{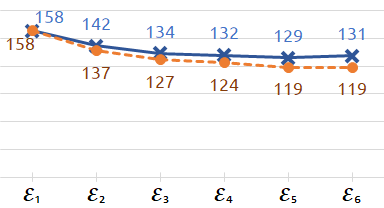}\vspace{-0.5em}
        \caption{\Sketch~\citep{cav15/jeon/adaptive, sttt13/solar-lezama/program}}
    \end{subfigure}%
    \begin{subfigure}{0.333\textwidth}%
        \includegraphics[width=\linewidth]{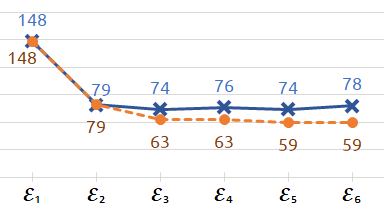}\vspace{-0.5em}
        \caption{\EUSolver~\citep{tacas17/alur/scaling}}
    \end{subfigure}
    \captionsetup{skip=1em}
    \caption{The number of failures on increasing grammar expressiveness,
             for state-of-the-art SyGuS tools, with and without the \PLearn framework (\cref{algo:plearn})\vspace{-0.2em}}
    \label{fig:ideal-failures-plot}
\end{figure}

\Cref{fig:ideal-failures-plot} summarizes our findings.
Without \PLearn the number of failures initially decreases and then increases across all solvers,
as grammar expressiveness increases.
However, with \PLearn the tools incur fewer failures at a given level of expressiveness, and there is a trend of \emph{decreased} failures with increased expressiveness.  
Thus, we have demonstrated that \PLearn is an effective measure to mitigate overfitting in SyGuS tools
and significantly improve their performance.

\vspace{-2em}\subsection{Performance of Hybrid Enumeration}
\label{sec:evaluation.he-loopinvgen}

\vspace{-0.3em}To evaluate the performance of hybrid enumeration,
we augment an existing synthesis engine with \HEAlgo(\cref{algo:hybrid-enum}).
We modify our \LoopInvGen tool~\citep{pldi16/padhi/data},
which is the best-performing SyGuS synthesizer from \cref{fig:ideal-failures-plot}.
Internally, \LoopInvGen leverages \textsc{Escher}~\citep{cav13/albarghouthi/escher},
an enumerative synthesizer, which we replace with \HEAlgo.
We make no other changes to \LoopInvGen.
We evaluate the performance and resource usage of this solver, \HELoopInvGen,
relative to the original \LoopInvGen with and without \PLearn (\cref{algo:plearn}).

\begin{figure}[!t]%
    \begin{subfigure}{0.66\textwidth}%
        \hspace{-0.25em}\vspace{-0.125em}%
        \includegraphics[width=0.975\linewidth]{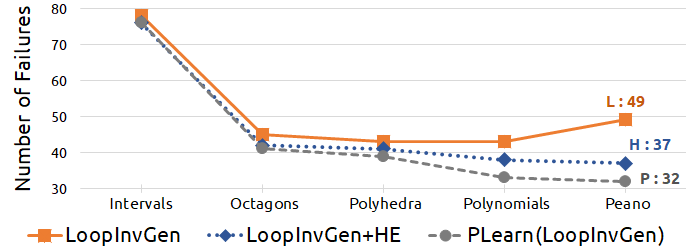}
        \captionsetup{skip=0.5em}
        \caption{Failures on increasing grammar expressiveness}
        \label{fig:lig-failures-plot}
    \end{subfigure}\hfill%
    \begin{subfigure}{0.325\textwidth}%
        \relscale{0.725}\def\arraystretch{1.15}%
        \begin{tabularx}{\linewidth}{rCC}
            \toprule
                {\relscale{0.95} \textbf{Grammar}\,} &
                \textbf{M}$\big[\!\frac{\textbf{\totalcost{1.4}[P]}}{\textbf{\totalcost{1.4}[H]}}\!\big]$ &
                \textbf{M}$\big[\!\frac{\textbf{\totalcost{1.4}[H]}}{\textbf{\totalcost{1.4}[L]}}\!\big]$ \\
            \midrule
                {\relscale{0.95} \Equality\,}   & {\relscale{1.15} $1.00$} & {\relscale{1.15} $1.00$} \\
                {\relscale{0.95} \NoArith\,}    & {\relscale{1.15} $1.91$} & {\relscale{1.15} $1.04$} \\
                {\relscale{0.95} \NoMult\,}     & {\relscale{1.15} $2.84$} & {\relscale{1.15} $1.03$} \\
                {\relscale{0.95} \LIA\,}        & {\relscale{1.15} $3.72$} & {\relscale{1.15} $1.01$} \\
                {\relscale{0.95} \NLMult\,}     & {\relscale{1.15} $4.62$} & {\relscale{1.15} $1.00$} \\
                {\relscale{0.95} \NIA\,}        & {\relscale{1.15} $5.49$} & {\relscale{1.15} $0.97$} \\
            \bottomrule
        \end{tabularx}
        \captionsetup{skip=0.7em}
        \caption{Median(\textbf{M}) overhead}
        \label{fig:lig-cpu-time}
    \end{subfigure}
    \captionsetup{skip=0.75em}
    \caption{\textbf{L}\,\textdblhyphen\,\LoopInvGen,
             \textbf{H}\,\textdblhyphen\,\HELoopInvGen,
             \textbf{P}\,\textdblhyphen\,\PLearn\!(\LoopInvGen).
             \textbf{H} is not only significantly robust against increasing grammar expressiveness,
             but it also has a smaller total-time cost (\texttau) than \textbf{P} and a negligible overhead over \textbf{L}.\vspace{-0.25em}}
    \label{fig:lig-he-vs-plearn}
\end{figure}

\vspace{-0.7em}\paragraph{Performance.}
In \cref{fig:lig-failures-plot}, we show the number of failures across our six grammars
for \LoopInvGen, \HELoopInvGen and \LoopInvGen with \PLearn, over our $180$ benchmarks.
\HELoopInvGen has a significantly lower failure rate than \LoopInvGen,
and the number of failures decreases with grammar expressiveness.
Thus, hybrid enumeration is a good proxy for \PLearn.

\vspace{-0.7em}\paragraph{Resource Usage.}
To estimate how computationally expensive each solver is,
we compare their \emph{total-time cost} (\totalcost{1.15}).
Since \LoopInvGen and \HELoopInvGen are single-threaded,
for them we simply use the wall-clock time for synthesis as the total-time cost.
However, for \PLearn with $p$ parallel instances of \LoopInvGen,
we consider the total-time cost as $p$ times the wall-clock time for synthesis.

In \cref{fig:lig-cpu-time}, we show the median overhead (ratio of \totalcost{1.15})
incurred by \PLearn over \HELoopInvGen and \HELoopInvGen over \LoopInvGen,
at various expressiveness levels.
As we move to grammars of increasing expressiveness,
the total-time cost of \PLearn increases significantly,
while the total-time cost of \HELoopInvGen essentially matches that of \LoopInvGen.

\vspace{-0.75em}\subsection{Competition Performance}
\label{sec:evaluation.competition}

\vspace{-0.25em}Finally, we evaluate the performance of \HELoopInvGen on the benchmarks from the \texttt{Inv} track
of the 2018 SyGuS competition~\citep{corr18/alur/sygus-comp},
against the official winning solver, which we denote \textsc{LIG}~\citep{corr18/padhi/loopinvgen} ---
a version of \LoopInvGen~\citep{pldi16/padhi/data} that has been extensively tuned for this track.
In the competition, there are some invariant-synthesis problems
where the postcondition itself is a satisfying expression.
\textsc{LIG} starts with the postcondition as the first candidate
and is extremely fast on such programs.
For a fair comparison, we added this heuristic to \HELoopInvGen as well.
No other change was made to \HELoopInvGen.

\LoopInvGen solves $115$ benchmarks in a total of $2191$\,seconds
whereas \HELoopInvGen solves $117$ benchmarks in $429$\,seconds,
for a mean speedup of over $5 \times$.
Moreover, no entrants to the competition could solve~\citep{corr18/alur/sygus-comp} the two additional benchmarks
(\texttt{\small gcnr\_tacas08} and \texttt{\small fib\_20}) that \HELoopInvGen solves.
\section{Related Work}
\label{sec:related-work}

\vspace{-0.125em}The most closely related work to ours investigates overfitting for verification tools~\cite{popl14/sharma/bias-variance}.
%They use cross-validation to combat overfitting in tuning a specific hyperparameter of a verifier.
Our work differs from theirs in several respects.
First, we address the problem of overfitting in CEGIS-based synthesis.
Second, we formally define overfitting and prove that all synthesizers must suffer from it,
whereas they only observe overfitting empirically.
%Third, our measures to mitigate overfitting are completely different,
%due to our focus on synthesis rather than verification.
Third, while they use cross-validation to combat overfitting in tuning a specific hyperparameter of a verifier,
our approach is to search for solutions at different expressiveness levels.

The general problem of efficiently searching a large space of programs for synthesis has been explored in prior work.
\citet{pldi18/lee/euphony} use a probabilistic model, learned from known solutions to synthesis problems,
to enumerate programs in order of their likelihood.
Other approaches employ type-based pruning of large search spaces~\citep{pldi16/polikarpova/program, pldi15/osera/type}.
These techniques are orthogonal to, and may be combined with,
our approach of exploring grammar subsets.

Our results are widely applicable to existing SyGuS tools, but some tools fall outside our purview.
For instance, in programming-by-example (PBE) systems~\citep[\S 7]{ftpl17/gulwani/ps},
the specification consists of a set of input-output examples.
Since any program that meets the given examples is a valid satisfying expression,
our notion of overfitting does not apply to such tools.
However in a recent work, \citet{pacmpl18/inala/webrelate} show that incrementally increasing expressiveness can also aid PBE systems.
They report that searching within increasingly expressive grammar subsets
requires significantly fewer examples to find expressions that generalize better over unseen data.
Other instances where the synthesizers can have a free lunch, \ie always generate a solution with a small number of counterexamples,
include systems that use grammars with limited expressiveness~\citep{icse10/jha/dit,esop13/sharma/algebraic,pldi12/godefroid/automated}.

Our paper falls in the category of formal results about SyGuS.
In one such result, \citet{acta17/jha/theory} analyze the effects of different kinds of counterexamples and
of providing bounded versus unbounded memory to learners.
Notably, they do not consider variations in ``concept classes'' or
``program templates,'' which are precisely the focus of our study.
Therefore, our results are complementary:
we treat counterexamples and learners as opaque and instead focus on grammars.
\vspace{-0.125em}\section{Conclusion}
\label{sec:conclusion}

\vspace{-0.125em}Program synthesis is a vibrant research area;
new and better synthesizers are being built each year.
This paper investigates a general issue
that affects all CEGIS-based SyGuS tools.
We recognize the problem of overfitting, formalize it,
and identify the conditions under which it must occur.
Furthermore, we provide mitigating measures for overfitting
that significantly improve the existing tools.

{\small\vspace{0.5em}%
\paragraph{\textnormal{\textbf{\ackname}}}

We thank Guy Van den Broeck and the anonymous reviewers for helpful feedback for improving this work,
and the organizers of the SyGuS competition for making the tools and benchmarks publicly available.

This work was supported in part by the National Science Foundation (NSF)
under grants CCF-1527923 and CCF-1837129.
The lead author was also supported by an internship and a PhD Fellowship from Microsoft Research.%
\par}

\renewcommand{\bibsection}{\section*{References}}
\bibliographystyle{splncsnat}
\begingroup
  \small
  \bibliography{paper}
\endgroup

\newpage
\appendix
\section{Proofs of Theorems}
\label{sec:appendix.proofs}

\subsection{No-Free-Lunch Theorems (\cref{sec:formalization.no-free-lunch})}
\label{sec:appendix.nfl-proofs}

\setcounter{theorem}{0}
\begin{theorem}[NFL in CEGIS-based SyGuS on Finite Sets]\label{thm:nfl-cegis-finite-appendix}%
    Let $X$ and $Y$ be two arbitrary finite sets,
    \theory be a theory that supports equality,
    \grammar be a grammar over \theory,
    and $m$ be an integer such that $0 \leq m < |X|$.
    Then, either:% \footnote{%
    %     We use $\permute{\mathlarger n}{\,\mathlarger r}$ to denote the number of \emph{permutations},
    %     \ie different ways of selecting an ordered subset of $r$ distinct elements from a set of $n$ distinct elements.}
    \begin{itemize}[itemsep=0.125em,topsep=0.125em]
        \item \grammar is not $k$-expressive for any $k > \sum_{i \,=\, 0}^{m}$ {\larger $\frac{|X|\bm{!}\:\;|Y|^i}{(|X| \,-\, i)\bm{!}}$}, or
        \item for every CEGIS-based learner \learner,
              there exists a satisfiable SyGuS problem $\query = \sygustuple$
              such that \query is not $m$-learnable by \learner.
              Moreover, there exists a different CEGIS-based learner for which \query is $m$-learnable.
    \end{itemize}
\end{theorem}

\begin{proof}%
    First, note that there are $t = \sum_{i \,=\, 0}^{m} \frac{|X|\bm{!}\:\;|Y|^i}{(|X| \,-\, i)\bm{!}}$
    distinct traces (sequences of counterexamples) of length at most $m$ over $X$ and $Y$.
    Now, consider some CEGIS-based learner \learner, and suppose \grammar is $k$-expressive for some $k > t$.
    Then, since the learner can deterministically choose at most $t$ candidates for the $t$ traces,
    there must be at least one function $f$ that is expressible in \grammar,
    but does not appear in the trace of $\learner^{\oracle[]}\!(m, \grammar)$ for any oracle \oracle[].

    Let $e$ be an expression in \grammar that implements the function $f$.
    Then, we can define the specification $\spec(f, x) \defeq f(x) = e(x)$ and the SyGuS problem $\query = \sygustuple$.
    By construction, \query is satisfiable since $e \models \query$,
    but we have that $\learner^{\oracle[]}\!(m, \grammar) \not\models \query$ for all oracles \oracle[].
    So, by \cref{def:learnability}, we have that \query is not $m$-learnable by \learner.

    However, we can construct a learner $\learner'$ such that \query is $m$-learnable by $\learner'$.
    We construct $\learner'$ such that $\learner'$ always produces $e$ as its first candidate expression for any trace.
    The result then follows by \cref{def:learnability}.
    \qed
\end{proof}

\setcounter{theorem}{1}
\begin{theorem}[NFL in CEGIS-based SyGuS on Countably Infinite Sets]\label{thm:nfl-cegis-infinite-appendix}%
    \!\!Let $X$ be an arbitrary countably infinite set,
    $Y$ be an arbitrary finite or countably infinite set,
    \theory be a theory that supports equality,
    \grammar be a grammar over \theory,
    and $m$ be an integer such that $m \geq 0$.
    Then, either:
    \begin{itemize}[itemsep=0.125em,topsep=0.125em]
        \item \grammar is not $k$-expressive for any $k > \aleph_0$,
              where $\aleph_0 \defeq | \naturals |$, or
        \item for every CEGIS-based learner \learner,
              there exists a satisfiable SyGuS problem $\query = \sygustuple$
              such that \query is not $m$-learnable by \learner.
              Moreover, there exists a different CEGIS-based learner for which \query is $m$-learnable.
    \end{itemize}
\end{theorem}

\begin{proof}%
    Consider some CEGIS-based learner \learner, and suppose \grammar is $k$-expressive for some $k > \aleph_0$.
    Note that there are $\sum_{i \,=\, 0}^{m} \frac{|X|\bm{!}\:\;|Y|^i}{(|X| \,-\, i)\bm{!}}$
    distinct traces of length at most $m$ over $X$ and $Y$.
    Let us overapproximate each $\frac{|X|\bm{!}\:\;|Y|^i}{(|X| \,-\, i)\bm{!}}$ as $(|X| \, |Y|)^m$,
    and thus the number of distinct traces as $(m+1) \, (|X| \, |Y|)^m$.
    We have two cases for $Y$:
    \begin{enumerate}[itemsep=0.125em,topsep=0.125em]
        \item $Y$ is finite \ie $|X| = \aleph_0$ and $|Y| < \aleph_0$.
              Then, the number of distinct traces is at most
              $(m + 1) \, (|X| \, |Y|)^m = (\aleph_0 \, |Y|)^m = \aleph_0$. Or,
        \item $Y$ is countably infinite \ie $|X| = |Y| = \aleph_0$.
              Then, the number of distinct traces is at most
              $(m + 1) \, (|X| \, |Y|)^m = (\aleph_0 \, \aleph_0)^m = \aleph_0$.
    \end{enumerate}

    Thus, the number of distinct traces is at most $\aleph_0$, \ie countably infinite.
    Since the number of distinct functions $k > \aleph_0$,
    the claim follows using a construction similar to the proof of \cref{thm:nfl-cegis-finite-appendix}.
    \qed
\end{proof}
\subsection{Overfitting Theorems (\cref{sec:formalization.overfitting})}
\label{sec:appendix.overfitting-proofs}

\setcounter{theorem}{2}
\begin{theorem}[Overfitting in SyGuS on Finite Sets]\label{thm:overfitting-finite-appendix}%
    Let $X$ and $Y$ be two arbitrary finite sets,
    $m$ be an integer such that $0 \leq m < |X|$,
    \theory be a theory that supports equality, and
    \grammar be a $k$-expressive grammar over \theory
    for some $k >$ {\larger $\frac{|X|\bm{!}\:\;|Y|^m}{m\bm{!}\;(|X|\,-\,m)\bm{!}}$}.
    Then, there exists a satisfiable SyGuS problem $\query = \sygustuple$
    such that $\doo(\query, Z) > 0$, for every set $Z$ of $m$ IO examples for \spec.
\end{theorem}

\begin{proof}%
    First, note that there are $t = \frac{|X|\bm{!}\:\;|Y|^m}{m\bm{!}\;(|X|\,-\,m)\bm{!}}$
    distinct ways of constructing a set of $m$ IO examples, over $X$ and $Y$.
    Now, suppose \grammar is $k$-expressive for some $k > t$.
    Then, there must be at least one function $f$ that is expressible in \grammar,
    but every set of $m$ IO examples that $f$ is consistent with is also satisfied by some other expressible function.

    Let $e$ be an expression in \grammar that implements the function $f$.
    Then, we can define the specification $\spec(f, x) \defeq f(x) = e(x)$
    and the SyGuS problem $\query = \sygustuple$.
    The claim then immediately follows from \cref{defn:overfitting}.
    \qed
\end{proof}

\setcounter{theorem}{3}
\begin{theorem}[Overfitting in SyGuS on Countably Infinite Sets]%
    Let $X$ be an arbitrary countably infinite set,
    $Y$ be an arbitrary finite or countably infinite set,
    \theory be a theory that supports equality, and
    \grammar be a $k$-expressive grammar over \theory
    for some $k > \aleph_0$.
    Then, there exists a satisfiable SyGuS problem $\query = \sygustuple$
    such that $\doo(\query, Z) > 0$, for every set $Z$ of $m$ IO examples for \spec.
\end{theorem}

\begin{proof}%
    Let us overapproximate the number of distinct ways of constructing a set of $m$ IO examples,
    $\frac{|X|\bm{!}\:\;|Y|^m}{m\bm{!}\;(|X|\,-\,m)\bm{!}}$ as $(|X| \, |Y|)^m$.
    Using cardinal arithmetic, as shown in the the proof of \cref{thm:nfl-cegis-infinite-appendix},
    this number is always at most $\aleph_0$.
    Then the claim follows using a construction similar to the proof of \cref{thm:overfitting-finite-appendix}.
    \qed
\end{proof}

\setcounter{theorem}{4}
\begin{theorem}[Overfitting Increases with Expressiveness]%
    Let $X$ and $Y$ be two arbitrary sets,
    \theory be an arbitrary theory,
    $\grammar_1$ and $\grammar_2$ be grammars over \theory
    such that $\grammar_1 \subseteq \grammar_2$,
    \spec be an arbitrary specification over \theory and a function symbol $f \colon X \to Y$,
    and $Z$ be a set of IO examples for \spec.
    Then, we have
    \begin{center}$
        \doo\big(\sygustuple[f][X][Y][\spec][\grammar_1], Z\big)
        \;\leq\;
        \doo\big(\sygustuple[f][X][Y][\spec][\grammar_2], Z\big)
    $\end{center}
\end{theorem}

\begin{proof}%
    If $\grammar_1 \subseteq \grammar_2$, then for any set $Z \subseteq X \times Y$ of IO examples, we have
    \begin{center}$
        \left\{ e \in \grammar_1 \,\mid\, \forall \io \in Z \colon e(x) = y \right\}
        \;\subseteq\;
        \left\{ e \in \grammar_2 \,\mid\, \forall \io \in Z \colon e(x) = y \right\}
    $\end{center}
    The claim immediately follows from this observation and \cref{defn:overfitting}.
    \qed
 \end{proof}
\subsection{Properties of Hybrid Enumeration (\cref{sec:mitigation.hybrid})}
\label{sec:appendix.he-proofs}

\begin{lemma}\label{lem:component-subset}%
    Let $\grammar_1$ and $\grammar_2$ be two arbitrary component-based grammars.
    Then, if $\grammar_1 \subseteq \grammar_2$,
    it must also be the case that $\basecomps{\grammar_1} \subseteq\, \basecomps{\grammar_2}$,
    where $\basecomps{\grammar_i}$ denotes the set of all components appearing in $\grammar_i$.
\end{lemma}

\begin{proof}
    Let $\components_1 = \basecomps{\grammar_1}$, $\components_2 = \basecomps{\grammar_2}$, and $\grammar_1 \subseteq \grammar_2$.
    Suppose $\components_1 \not\subseteq \components_2$.
    Then, there must be at least one component $c$ such that $c \in \components_1 \setminus \components_2$.
    By definition of $\basecomps{\grammar_1}$,
    the component $c$ must appear in at least one expression $e \in \grammar_1$.
    However, since $c \not\in \components_2$, it must be the case that $e \not\in \grammar_2$,
    thus contradicting $\grammar_1 \subseteq \grammar_2$.
    Hence, our assumption $\components_1 \not\subseteq \components_2$ must be false.
    \qed
\end{proof}

\setcounter{theorem}{6}
\begin{theorem}%
    Given component-based grammars \grammars{p},
    the following strict partial order $\,\order_\mathlarger{\mkern2mu\ast}$ on $\grammars{p} \times \naturals$
    is a well order
    \begin{center}\vspace{-0.0625em}$
        \forall\, \grammar_a, \grammar_b \in \grammars{p} :\enskip
            \forall\, m, n \in \naturals :\enskip
                (\grammar_a, m) \order_\mathlarger{\mkern2mu\ast} (\grammar_b, n)
                \iff
                |\mkern2mu\components_a\mkern1mu|^m < |\mkern2mu\components_b\mkern1mu|^n
    $\vspace{-0.0625em}\end{center}
    where $\components_i = \basecomps{\grammar_i}$ denotes the set of all components appearing in $\grammar_i$.
\end{theorem}
\begin{proof}%
    Let $\grammar_a$ and $\grammar_b$ be two component-based grammars in \grammars{p}.
    By \cref{lem:component-subset},
    we have that $\grammar_a \subseteq \grammar_b \implies \basecomps{\grammar_1} \subseteq \basecomps{\grammar_2}$.
    The claim then immediately follows from \cref{def:well-order}.
    \qed
\end{proof}

\begin{definition}[$\bm{j^k}$-Uniqueness]\label{def:j-k-uniqueness}%
    Given grammars $\grammar_1 \subseteq \cdots \subseteq \grammar_p$,
    we say that an expression $e$ of size $k$ is $j^k$-unique with respect to \grammars{p}
    if it is contained in $\grammar_j$ but not in $\grammar_{(j-1)}$.
    We define $\maxunique$ as the maximal such set of expressions, \ie
    \begin{center}\vspace{-0.0625em}$
        \maxunique \;\defeq\;
            \left\{
                e \in \grammar_j
                    \;\mid\;
                        \text{\sf size}(e) = k
                    \;\bwedge\;
                        e \not\in \grammar_{(j-1)}
            \right\}
    $\vspace{-0.0625em}\end{center}
\end{definition}

\begin{lemma}\label{lem:last-arg}%
    Let $\grammar_1 \subseteq \cdots \subseteq \grammar_p$ be $p$ component-based grammars.
    Then, for any expression $o(e_1, \ldots, e_a) \in \maxunique$,
    if the operator $o$ belongs to $\operators{\grammar_q}$ such that $q < j$,
    at least one argument must belong to $\grammar_j$ but not $\grammar_{(j-1)}$, \ie
    \begin{center}\vspace{-0.0625em}$
        o \in \operators{\grammar_q} \,\bwedge\, q < j
        \;\implies\;
        \exists\, e \in \expressions{a} \colon e \in \grammar_j \bwedge e \not\in \grammar_{(j-1)}
    $\vspace{-0.0625em}\end{center}
\end{lemma}

\begin{proof}
    Consider an arbitrary expression $e_\ast = o(e_1,\ldots,e_a) \in \maxunique$
    such that $o \in \operators{\grammar_q} \;\bwedge\; q < j$.
    Suppose $\left[\forall\, e \in \expressions{a} \colon e \not\in \grammar_j \,\bvee\, e \in \grammar_{(j-1)}\right]$.
    Then, for any argument subexpression $e$, we have the following three possibilities:
    \begin{itemize}[leftmargin=3em,itemsep=0.25em,topsep=0.3125em]
        \item[\ding{55}] $e \not\in \grammar_j \,\wedge\, e \in \grammar_{(j-1)}$ is impossible since $\grammar_{(j-1)} \subseteq \grammar_j$.
        \item[\ding{55}] $e \not\in \grammar_j \,\wedge\, e \not\in \grammar_{(j-1)}$ is also impossible, by \cref{def:component-based-grammar},
                         due to the closure property of component-based grammars.
        \item[\ding{55}] $e \in \grammar_j \,\wedge\, e \in \grammar_{(j-1)}$ must be false for at least one argument subexpression.
                         Otherwise, since $o \in \operators{\grammar_{(j-1)}}$
                         and $\grammar_{(j-1)}$ is closed under operator application by \cref{def:component-based-grammar},
                         $e_\ast \in \grammar_{(j-1)}$ must be true.
                         However, by \cref{def:j-k-uniqueness}, we have that $e_\ast \in\, \maxunique \implies e_\ast \not\in \grammar_{(j-1)}$.
    \end{itemize}
    Therefore, our assumption $\left[\forall\, e \in \expressions{a} \colon e \not\in \grammar_j \,\bvee\, e \in \grammar_{(j-1)}\right]$ must be false.
    \qed
\end{proof}

\begin{lemma}\label{lem:divide}%
    Let $\grammar_0 = \{\}$ and $\grammar_1 \subseteq \cdots \subseteq \grammar_p$ be $p$ component-based grammars.
    Then, for any $l \geq 1$ and any operator $o \in \operators{\grammar_l} \setminus \operators{\grammar_{l-1}}$ of arity $a$,
    $\textnormal{\textsc{Divide}}(a, k-1, l, j, \langle\rangle)$ generates the following set $L$ of
    all possible distinct locations for selecting the arguments for $o$ such that $o(e_1, \ldots, e_a) \in \maxunique$:
    \begin{center}\vspace{-0.125em}$
        \begin{array}{rcl}
            L = \Big\{
                \!\big\langle (j_1, k_1), \ldots, (j_a, k_a) \big\rangle
                    & \scalebox{1.45}{$\mid$} &
                o(e_1, \ldots, e_a) \in \maxunique \\[0.25em]
                    & \bwedge &
                      \forall\, 1 \leq i \leq a \colon e_i \in \maxunique[\grammars{p}][j_i][k_i]
                \Big\}
        \end{array}
    $\end{center}
\end{lemma}

\begin{proof}%
    In lines \range{7}{9} of \cref{algo:divide-algo},
    the top-level $\textsc{Divide}(a, k-1, l, j, \langle\rangle)$ call first recursively creates a call tree of height $a-1$
    such that the accumulator $\alpha$ at each leaf node contains the locations for selecting the last $a-1$ arguments from.
    Since $u$ in line $7$ ranges over $\{1, \ldots, j\}$ and $v$ in line $8$ ranges over $\{1, \ldots, (k-2)\}$,
    the call tree must be exhaustive by construction.
    Concretely, the values of $\alpha$ at the the leaf nodes must capture every possible sequence of $a-1$ locations,
    $\left\langle (j_1, k_1), \ldots, (j_{(a-1)}, k_{(a-1)}) \right\rangle$,
    such that $k_1 + \cdots + k_{(a-1)} \leq k-2$.

    Finally at the leaf nodes of the call tree, lines \range{4}{5} are triggered to select locations for the first argument.
    The na\"{i}ve approach of simply assigning the remaining size to each grammar in \grammars{j} would be exhaustive,
    but may lead to enumerating other expressions $o(e_1, \ldots, e_a) \not\in \maxunique$ when $l < j$.
    Therefore, we check if $l < j$ and no location $(x, y)$ in $\alpha$ satisfies $x = j$,
    in which case we assign the remaining size to only $\grammar_j$ in line $5$.
    \cref{lem:last-arg} shows that this check is sufficient to guarantee that we only enumerate expressions in \maxunique.
    \qed
\end{proof}

\setcounter{theorem}{7}
\begin{theorem}[\textnormal{\HEAlgo} is Complete up to Size $\bm{q}$]\label{thm:he-completeness-appendix}%
    Given a SyGuS problem $\query = \sygustuple$,
    let \grammars{p} be component-based grammars over theory \theory
    such that $\grammar_1 \subset \cdots \subset \grammar_p = \grammar$,
    $\order$ be a well order on $\grammars{p} \times \naturals$,
    and $q \geq 0$ be an upper bound on size of expressions.
    Then, \textnormal{\HEAlgo}$(\query, \grammars{p}, \order, q)$
    will eventually find a satisfying expression
    if there exists one with size $\leq q$.
\end{theorem}

\begin{proof}%
    First, we observe that every expression $e \in \grammar$
    must belong to \emph{some} maximal set of $j^k$-unique expressions with respect to \grammars{p}:
    \begin{center}$
        \forall\, e \in \grammar \colon\; \exists\, j \in \{1,\ldots,p\} \colon\; \exists\, k \in \{1,\ldots,q\} \colon\; e \in \maxunique
    $\end{center}

    We show that $C[j, k]$ in \HEAlgo (\cref{algo:hybrid-enum}) stores \maxunique,
    into various $C[j, k][\tau]$ lists based on the expression type $\tau$.
    Since \HEAlgo computes $C[j, k]$ for each $j \in \{1,\ldots,p\}$ and each $k \in \{1,\ldots,q\}$,
    it must enumerate every expression in \grammar with size at most $q$,
    and thus eventually find $e$.

    The base cases $C[i,1] = \maxunique[\grammars{p}][i][1]$ are straightforward.
    The inductive case follows from \cref{lem:divide}.
    For each $(j, k) \in \{1,\ldots,p\} \times \{1,\ldots,q\}$ and each operator in \grammars{j},
    we invoke \textsc{Divide} (\cref{algo:divide-algo}) to generate all possible locations for the operator's arguments
    such that the final expression is contained in \maxunique.
    Lines \range{16}{20} in \HEAlgo then populate $C[j, k]$ as $\maxunique$
    by applying the operator to subexpressions of appropriate types drawn from these locations.
    \qed
\end{proof}

\begin{lemma}\label{lem:unique-j-k-disjointness}%
    Given grammars $\grammar_1 \subseteq \cdots \subseteq \grammar_p$,
    for any distinct pairs $(j, k)$ and $(j', k')$
    the sets $\maxunique$ and $\maxunique[\grammars{p}][j'][k']$ must be disjoint, \ie
    \begin{center}$
        \forall\ j,k,j',k' \colon\; j \neq j' \bvee k \neq k' \;\implies\; \maxunique \,\cap\, \maxunique[\grammars{p}][j'][k'] = \{\}
    $\end{center}
\end{lemma}

\begin{proof}%
    When $k \neq k'$, it is straightforward to show that $\maxunique \,\cap\, \maxunique[\grammars{p}][j'][k'] = \{\}$,
    since an expression cannot be of size $k$ and $k'$ at the same time.

    We now prove the claim for the case when $j \neq j'$ by contradiction.
    Suppose there exists an expression $e \in \maxunique \,\cap\, \maxunique[\grammars{p}][j'][k]$.
    Without loss of generality, assume $j > j'$, and therefore $\grammar_j \supseteq \grammar_{j'}$.
    But then, by \cref{def:j-k-uniqueness},
    it must be the case that $e \not\in \grammar_{j'}$ and thus $e \not\in \maxunique[\grammars{p}][j'][k]$.
    Therefore, our assumption that $\maxunique \,\cap\, \maxunique[\grammars{p}][j'][k] \neq \{\}$ must be false.
    \qed
\end{proof}

\setcounter{theorem}{8}
\begin{theorem}[\textnormal{\HEAlgo} is Efficient]%
    Given a SyGuS problem $\query = \sygustuple$,
    let \grammars{p} be component-based grammars over theory \theory
    such that $\grammar_1 \subset \cdots \subset \grammar_p \subseteq \grammar$,
    $\order$ be a well order on $\grammars{p} \times \naturals$,
    and $q \geq 0$ be an upper bound on size of expressions.
    Then, \textnormal{\HEAlgo}$(\query, \grammars{p}, \order, q)$
    will enumerate each distinct expression at most once.
\end{theorem}

\begin{proof}%
    As shown in the proof of \cref{thm:he-completeness-appendix},
    $C[j,k]$ in \HEAlgo (\cref{algo:hybrid-enum}) stores \maxunique.
    Then, by \cref{lem:unique-j-k-disjointness}, we immediately have that all pairs $C[j,k]$ and $C[j',k']$
    of synthesized expressions are disjoint when $j \neq j'$ or $k \neq k'$.

    Furthermore, although each $C[j, k]$ is implemented as a list,
    we show that any two expressions within any $C[j, k]$ list must be syntactically distinct.
    The base cases $C[i, 1]$ are straightforward.
    For the inductive case, observe that if each list $C[j_1, k_1], \ldots, C[j_a, k_a]$ only contains syntactically distinct expressions,
    then all tuples within $C[j_1, k_1] \times \cdots \times C[j_a, k_a]$ must also be distinct.
    Thus, if an operator $o$ with arity $a$ is applied to subexpressions drawn from the cross product,
    \ie $\langle e_1, \ldots, e_a \rangle \in C[j_1, k_1] \times \cdots \times C[j_a, k_a]$,
    then all resulting expressions of the form $o(e_1, \ldots, e_a)$ must be syntactically distinct.
    Thus, by structural induction, we have that in any list $C[j, k]$ all contained expressions are syntactically distinct.
    \qed
\end{proof}

\end{document}